\documentclass[conference]{IEEEtran}
\usepackage{filecontents,lipsum}
\usepackage[noadjust]{cite}
\usepackage{graphicx} 
\usepackage{float}
\usepackage{tikz}
\usetikzlibrary{arrows.meta,positioning,fit,shapes.geometric,calc}
\usepackage[numbers]{natbib}
\usepackage{todonotes}
\usepackage{array}
\usepackage{makecell}
\usepackage{booktabs}
\usepackage{amssymb}
\usepackage{pifont}
\usepackage{xcolor}
\usepackage{amsmath}
\usepackage{listings}
\usepackage{pdflscape}
\usepackage{longtable}
\usepackage{tabularx}
\usepackage{enumitem}
\usepackage{comment}
\usepackage{hyperref}
\usepackage{url}
\lstdefinestyle{mermaid}{language=,basicstyle=\ttfamily,breaklines=true}
\newcolumntype{P}[1]{>{\raggedright\arraybackslash}p{#1}}
\newcolumntype{Y}{>{\raggedright\arraybackslash}X}
\tikzset{
  diagramnode/.style={draw, rounded corners, align=center, font=\scriptsize, inner sep=4pt, minimum height=0.72cm},
  diagramhead/.style={diagramnode, fill=blue!8, very thick},
  diagramtask/.style={diagramnode, fill=green!8},
  diagramtool/.style={diagramnode, fill=orange!12},
  diagramrisk/.style={diagramnode, fill=red!8},
  diagramarrow/.style={-{Latex[length=2.2mm]}, thick},
  diagramdash/.style={-{Latex[length=2.2mm]}, thick, dashed}
}

\newcommand{\cyes}{\ding{51}}

\title{A Survey on LLM-Integrated Hardware Design Verification}

\author{
  \IEEEauthorblockN{
    Hao Zheng\IEEEauthorrefmark{1},
    Jaime Rafael Imperial\IEEEauthorrefmark{1},
    Bardia Nadimi\IEEEauthorrefmark{1},
    Xiangfei Kong\IEEEauthorrefmark{1}
  }
  \IEEEauthorblockA{
    \IEEEauthorrefmark{1}Bellini College of Cybersecurity, AI, and Computing, University of South Florida\\
    Tampa, FL, USA\\
    Email: \{haozheng, imperialj, bnadimi, xiangfeikong\}@usf.edu
  }
}

\date{October 2025}

\begin{document}

\maketitle

\thispagestyle{plain}
\pagestyle{plain}

\begin{abstract}
Large language models (LLMs) are increasingly being integrated into hardware verification to automate specification interpretation, verification-artifact generation, debugging, formal reasoning, and tool orchestration.
This survey provides a systematic review of LLM-assisted hardware functional verification across SystemVerilog assertion generation, stimulus and testbench generation, bug localization and design repair, model checking and equivalence checking, SAT/SMT optimization, and emerging agentic verification workflows.
We organize the literature by methodology, verification objective, tool interaction, benchmark, and evaluation criterion, and examine both inference-time techniques---including prompting, retrieval, structured reasoning, and agentic workflows---and training-time adaptation.
Across these areas, a common pattern emerges: LLMs are most effective as semantic reasoning, search, and orchestration components embedded within verification-aware workflows, while simulators, formal engines, coverage tools, and solvers provide executable feedback and correctness evidence.
However, tool acceptance alone does not establish verification correctness, since assertions, tests, repairs, or proofs may satisfy available checks without faithfully capturing the complete design intent.
We therefore identify semantic alignment between specifications and verification evidence, scalable integration with deterministic tools, generalization to unseen designs, and rigorous evaluation of correctness, cost, robustness, and human effort as key challenges.
Finally, we discuss emerging directions toward specification-centered, neuro-symbolic, and persistent agentic verification systems that combine LLM flexibility with independently checkable verification evidence.
\end{abstract}


\section{Introduction}

Computing systems have become indispensable to modern society, supporting a wide range of applications with increasingly stringent requirements for functionality, performance, reliability, safety, security, and energy efficiency. 
At the hardware level, System-on-Chip (SoC) designs have become a dominant
architecture, with modern chips integrating tens to hundreds of billions of
transistors and increasingly heterogeneous and tightly coupled functional
blocks \cite{apple:m2ultra:2023,nvidia:blackwell:2024}. 
This growing hardware complexity, coupled with society's increasing reliance on computing systems in safety-, security-, and mission-critical applications, makes ensuring hardware correctness and dependability increasingly challenging and essential. Consequently, rigorous hardware verification is a fundamental requirement for identifying design errors and establishing confidence that hardware implementations conform to their intended specifications before deployment.

Consequently, hardware verification has become one of the most challenging, costly, and resource-intensive stages of the IC design process \cite{abvsurvey:witharana:2022}, aiming to establish that hardware implementations conform to their functional requirements and design intent. While verification spans both pre-silicon verification and post-silicon validation, pre-silicon verification is particularly critical because defects that escape into fabrication can require costly silicon re-spins, cause substantial schedule delays, and incur significant financial losses \cite{nodecost:moyer:2024}. Yet, growing hardware complexity has made pre-silicon verification increasingly difficult to scale. The 2024 Wilson Research Group Functional Verification Study~\cite{vstudy:foster:2024} reports that peak IC/ASIC verification teams average over 14 engineers, compared with approximately 12 design engineers, while design engineers spend about 50\% of their time on verification. Overall, verification is estimated to consume 60--70\% of IC development cost and schedule~\cite{Pan2025LLMSurvey,Yu2024MLFV}, highlighting the need for greater automation and productivity in modern hardware verification.

Large Language Models (LLMs)~\cite{GPT5OpenAI2025,Claude2025,Gemini2025} are increasingly used across the Electronic Design Automation (EDA) stack because they can interpret and generate diverse hardware artifacts, including specifications, RTL, assertions, testbenches, and tool logs. Their flexible natural-language interface enables reasoning across heterogeneous information and interaction with EDA tools through retrieval-augmented and agentic workflows~\cite{Pan2025Survey,Xu2025LLMEDA}. Consequently, LLMs have been explored for specification analysis~\cite{wang2025realbench,kumar2025enhancing}, RTL and hardware generation~\cite{liu2024rtlcoder,ho2025verilogcoder,zhao2025mage,nadimi2024multiexpert}, design-space exploration~\cite{sun2026edacopilot}, verification planning~\cite{kumar2025saarthi}, assertion and testbench generation~\cite{CorrectBench,verilogreader2024,assertllm:fang:2024,sangam:gupta:2025,assertionforge:bai:2025}, bug localization and repair~\cite{dbg_lik, rtlfixer}, and autonomous verification~\cite{Zang2026Agentic}. Recent approaches increasingly integrate LLMs with simulators, formal verification engines, synthesis tools, and external knowledge to ground generation and enable iterative refinement~\cite{ChipNeMo2024,Zang2026Agentic}. 
Despite substantial progress, significant challenges remain in ensuring correctness, mitigating hallucinations, preserving design intent, scaling to industrial-scale designs, and establishing rigorous evaluation methodologies tailored to hardware engineering tasks motivating systematic research into trustworthy and autonomous LLM-integrate hardware verification.

\subsection{Why a New Survey Focused on LLM-Driven Verification?}

The rapid adoption of Large Language Models (LLMs) has spurred a growing body of survey literature on their applications in Electronic Design Automation (EDA). 
Existing surveys consistently identify LLMs as a transformative technology across the chip design flow, but they differ substantially from the scope of this survey. 
Pan et al.~\cite{Pan2025Survey} provide a broad EDA-wide treatment spanning system-level design, RTL design, logic synthesis, physical design, and analog design, with verification appearing as one component of a much larger design-automation landscape. 
Xu et al.~\cite{Xu2025LLMEDA} similarly frame LLMs as a general technology across the chip flow, including specification optimization, HDL generation, testing, debugging, synthesis, and back-end task execution. 
Consequently, they relegate functional verification to brief sub-sections, missing technical nuances like Universal Verification Methodology (UVM) infrastructure composition.
Yang et al.\cite{Yang2025Verilog} narrow the focus to Verilog code generation, explicitly organizing the literature around models, datasets, metrics, generation techniques, and alignment, but their inclusion criteria exclude studies focused solely on repair or testing without generation. 
The recent Agentic EDA survey~\cite{Zang2026Agentic} targets autonomous orchestration of the RTL-to-GDSII flow and emphasizes perception--cognition--action loops, tool execution, and the sim-to-silicon gap rather than the detailed methodology of verification tasks. 
The work in~\cite{security2026ai} focuses primarily on vulnerability identification (e.g., Common Weakness Enumerations) and threat modeling. While they leverage testbenches and assertions, their core metric is security compliance rather than general functional coverage closure.
A verification-specific survey by Liu et al.~\cite{Liu2026Verification} is closest in spirit, but it is compact and organized as a high-level overview of assertion generation, testbench generation, debugging, and holistic verification frameworks. 
Consequently, these works do not provide comprehensive coverage of the rapidly expanding landscape of LLM-driven hardware functional verification, including emerging techniques for testbench generation, assertion generation, coverage closure, bug localization, debugging, design repair, verification agents, benchmark development, and evaluation methodologies. This gap has become increasingly pronounced as the field has grown rapidly since 2024, with dozens of verification-focused studies appearing in leading EDA venues. 
As a result, researchers and practitioners currently lack a unified and systematic understanding of how LLMs are being applied across the full verification workflow, the associated benchmark ecosystems, and the remaining technical challenges. 
Table~\ref{tab:survey_comparison} summarizes the structural coverage gaps of existing literature across the four pillars of the proposed survey paper.

\begin{table*}[t]
\centering
\caption{Comparison of existing surveys and the proposed survey. 
A checkmark (\cyes) indicates substantial coverage of a topic, 
while $\triangle$ denotes partial or high-level discussion.}
\label{tab:survey_comparison}
\renewcommand{\arraystretch}{1.15}
\resizebox{\textwidth}{!}{
\begin{tabular}{p{0.7in}p{0.5in}p{0.5in}p{0.5in}p{0.5in}p{0.3in}p{0.3in}p{0.5in}p{0.5in}p{5in}}
\hline
\textbf{Survey} &
\textbf{EDA-wide} &
\textbf{RTL Gen.} &
\textbf{SVA Gen.} &
\textbf{Testbench / Stimulus} &
\textbf{Debug} &
\textbf{Repair} &
\textbf{Benchmarks} &
\textbf{Verification-Centric} &
\textbf{Scope} \\
\hline

Pan et al.~\cite{Pan2025Survey} &
\cyes &
\cyes &
$\triangle$ &
$\triangle$ &
$\triangle$ &
$\triangle$ &
$\triangle$ &
$\times$ &
Broad survey of LLM applications across EDA workflow \\

Xu et al.~\cite{Xu2025LLMEDA} &
\cyes &
\cyes &
$\triangle$ &
$\triangle$ &
$\triangle$ &
$\triangle$ &
$\times$ &
$\times$ &
Overview of LLMs in design, testing, optimization, and automation \\

Yang et al.~\cite{Yang2025Verilog} &
$\times$ &
\cyes &
$\times$ &
$\times$ &
$\times$ &
$\times$ &
\cyes &
$\times$ &
Focused on Verilog/RTL code generation models, datasets, and evaluation \\

Zang et al.~\cite{Zang2026Agentic} &
\cyes &
\cyes &
$\times$ &
$\times$ &
$\triangle$ &
$\triangle$ &
$\times$ &
$\times$ &
Agentic EDA and autonomous RTL-to-GDSII workflows \\

Liu et al.~\cite{Liu2026Verification} &
$\times$ &
$\times$ &
\cyes &
\cyes &
\cyes &
$\triangle$ &
$\triangle$ &
\cyes &
Overview of LLM-assisted circuit verification \\

\hline

\textbf{This Survey} &
$\times$ &
$\triangle$ &
\cyes &
\cyes &
\cyes &
\cyes &
\cyes &
\cyes &
Comprehensive survey of LLMs for verification, including assertion generation, testbench/stimulus generation, bug localization, debugging, RTL repair, datasets, benchmarks, evaluation methodologies, tool integration, and open challenges \\

\hline
\end{tabular}
}
\end{table*}

This survey provides a comprehensive review of recent research on LLM-integrated hardware functional verification, a rapidly emerging research area with unique workflows, artifacts, evaluation criteria, and tool-integration requirements that are not adequately captured by broader LLM-for-EDA surveys. Unlike prior reviews that treat verification as a secondary component of the design flow, this survey systematically organizes the literature across the complete verification pipeline, including SystemVerilog Assertion (SVA) generation, testbench and stimulus generation,  bug localization, debugging, design repair, and autonomous verification agents. 
In addition, the survey provides a unified comparison of datasets, benchmarks, tools, and evaluation frameworks used throughout the field, identifies common limitations such as hallucination, tool orchestration, and scalability, and highlights emerging research directions including multimodal LLMs, agentic verification systems, and human-in-the-loop verification workflows. 
By treating hardware verification as an integrated feedback-driven process rather than a collection of isolated applications, this survey establishes a comprehensive reference for researchers and practitioners and provides a roadmap for advancing LLM-integrated verification from proof-of-concept demonstrations to dependable engineering practice.


\subsection{Paper Organization}

The organization of this paper is as follows. 
Section~\ref{sec:background} gives background necessary to the following sections.
Section~\ref{sec:sva-gen} surveys the latest research on SVA generations using LLMs.
Section~\ref{sec:tb-gen} surveys the latest research on stimulus and testbench generation usinig LLMs.
Section~\ref{sec:debug} surveys the latest research on bug localization and design repair using LLMs.
Section~\ref{sec:llm4fv} surveys the latest research on formal verification using LLMs.
Section~\ref{sec:conclusion} concludes the paper, offers our views of current challenges to LLM-based verification methods, and highlight promising future research directions.

\section{Background}
\label{sec:background}

\subsection{Overview of Hardware Design Flow}

\begin{figure}[t]
    \centering
    \includegraphics[width=1\linewidth]{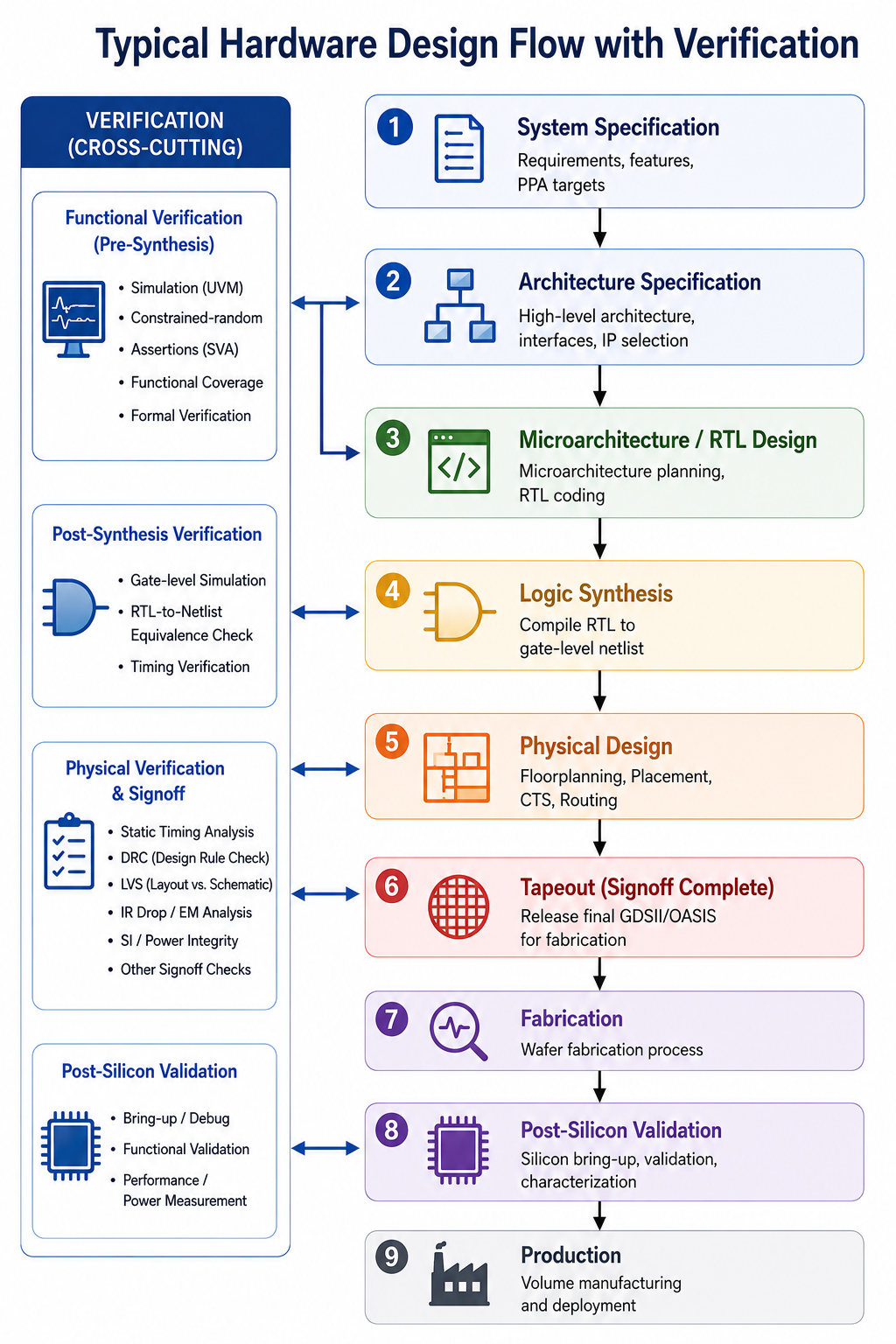}
    \caption{A simplified overview of hardware design flow.}
    \label{fig:design-flow}
\end{figure}

A typical hardware design flow, illustrated in Fig.~\ref{fig:design-flow}, progresses through increasingly concrete representations from system requirements to fabricated silicon. System and architectural specifications define functional requirements, interfaces, and performance, power, and area (PPA) objectives, which are refined into microarchitecture and synthesizable RTL. The RTL is transformed through logic synthesis into a gate-level netlist and subsequently through physical design, including floorplanning, placement, clock-tree synthesis, and routing, followed by physical verification and signoff before tapeout and fabrication. Fabricated devices then undergo post-silicon validation for functionality, performance, and power.

Verification is a cross-cutting activity throughout this flow rather than an isolated design stage. At the RTL and gate levels, functional verification employs simulation, constrained-random testing, assertions, functional coverage, formal methods, and RTL-to-netlist equivalence checking. Physical implementation is validated through static timing analysis, design-rule checking (DRC), layout-versus-schematic (LVS) checking, and power- and signal-integrity analysis, followed by post-silicon validation. Because verification can expose design defects or specification violations at any stage, the flow is inherently iterative, with verification results potentially triggering revisions to earlier design stages. This survey focuses on functional (logic) verification at the RTL and gate levels.

\subsection{Verification Concepts and Techniques}

Hardware functional verification aims to establish that an implementation conforms to its intended functional and temporal behavior. 
It primarily combines simulation-based verification, formal methods, and hardware-assisted techniques. Simulation remains the dominant approach, exercising the design with directed or constrained-random stimuli and using coverage metrics to guide verification progress; methodologies such as the Universal Verification Methodology (UVM) support reusable and modular verification environments~\cite{accellera2015uvm1}. Formal methods, including model checking and equivalence checking, mathematically establish specified properties or behavioral equivalence and can expose corner cases that are difficult to reach through simulation~\cite{gupta1992formal,clarke1999model,bmc:2004}. These techniques commonly rely on Boolean SAT~\cite{sat-minisat:2005,sat-CaDiCaL:2019} and Satisfiability Modulo Theories (SMT) solvers~\cite{smt-z3:2008,smt-cvc5:2022}, while emulation and FPGA-based prototyping accelerate the execution of large workloads that are impractical to explore through software simulation.

A typical functional-verification flow develops a testbench that generates stimulus, drives the design under test (DUT), monitors responses, and checks observed behavior against expected results. Verification engineers must also develop test scenarios, assertions, checkers, and coverage models, followed by iterative simulation, failure diagnosis, and regression testing. 
Coverage metrics help identify untested functionality and guide subsequent verification, but no single metric can establish verification completeness~\cite{accellera2015uvm1}. 
Moreover, verification remains difficult to scale because modern designs exhibit enormous input and state spaces, long temporal behaviors, and complex interactions among components and software. 
Simulation may fail to exercise difficult-to-reach behaviors, while formal methods face state-space explosion and scalability limitations~\cite{gupta1992formal,ieee2026formal}. 
Developing verification artifacts, diagnosing failures determining whether apparent coverage gaps represent genuine verification deficiencies also requires substantial hardware expertise and engineering effort. 
These challenges motivate increasing interest in automation, including LLM-based methods for generating verification artifacts, exploring design behavior, and assisting verification and debugging.

\subsection{Large Language Models}
\label{bg:llms}

Large Language Models (LLMs) are deep learning models trained on large-scale text and source-code corpora to learn general-purpose representations of language, programming constructs, and domain knowledge. Most modern LLMs are based on the Transformer architecture~\cite{vaswani2017attention}, whose self-attention mechanism enables effective modeling of long-range dependencies and highly parallelized training. LLMs are typically developed through large-scale pre-training followed by post-training, such as supervised fine-tuning and preference- or feedback-based optimization, which adapts general capabilities toward instruction following and task-specific behavior.

LLMs can be adapted to downstream tasks through prompting, including few-shot and reasoning-oriented strategies such as chain-of-thought prompting~\cite{wei2022chain}. These capabilities are particularly relevant to hardware verification, where an LLM may need to interpret specifications, identify relationships among signals and modules, generate assertions or test scenarios, and reason about verification failures. However, LLM-generated reasoning and artifacts are not correctness guarantees, motivating the use of external knowledge and deterministic verification mechanisms.

Retrieval-Augmented Generation (RAG)~\cite{lewis2020rag} supplements model knowledge with task-specific information retrieved at inference time. For hardware verification, retrieved context may include specifications, RTL modules, interface descriptions, protocol documentation, assertions, verification plans, or debugging records. LLMs can also interact with external tools, using approaches such as ReAct~\cite{yao2023react} and Toolformer~\cite{schick2023toolformer}, to invoke simulators, formal engines, compilers, coverage analyzers, or other EDA tools and use their outputs to refine subsequent actions. This tool-grounded interaction is particularly important for verification because deterministic tools can independently evaluate generated artifacts.

Agentic LLM systems extend these capabilities by decomposing complex tasks into multiple steps, maintaining intermediate state, invoking tools, and iteratively refining results; multi-agent frameworks can further assign specialized roles such as planning, generation, verification, and debugging~\cite{wu2023autogen}. These workflows, however, must manage the limited context available to each model invocation. Large RTL hierarchies, specifications, testbenches, waveforms, and tool logs can exceed practical context limits, motivating long-context models, retrieval, summarization, and structured representations. Together, reasoning, retrieval, tool use, and agentic orchestration provide the foundation for applying LLMs to hardware verification, while deterministic verification tools remain essential for validating their outputs.

\subsection{LLMs for EDA}
\label{bg:llm4eda}

The integration of Large Language Models (LLMs) into Electronic Design Automation (EDA) flows is enabling increasingly autonomous and machine-augmented hardware design workflows. Early efforts such as VeriGen demonstrated domain adaptation by fine-tuning open-source models on Verilog corpora [1]. Subsequent approaches have explored natural-language-driven RTL generation and iterative refinement, with frameworks such as AutoChip incorporating compiler and simulation feedback into multi-turn generation loops to identify and correct syntactic and functional errors [2], while RTLCoder demonstrated the effectiveness of domain-specific training data and lightweight open-source models on benchmarks such as RTLLM and VerilogEval [3].

Hardware verification is another major application area. LLM-based approaches have been developed to automate verification collateral generation, including testbenches through frameworks such as AutoBench [4], and assertion generation from natural-language design specifications through systems such as AssertLLM [5]. These methods reduce the manual effort required to translate design intent into executable verification artifacts while retaining conventional simulation and formal-verification tools as the final validation mechanisms.

Beyond RTL generation and verification, LLMs are being applied across the EDA optimization stack. At the HLS level, approaches such as C2HLSC and HLSPilot use LLMs to transform and optimize HLS code [6], [7]. At the physical-design level, agent-based frameworks such as ChatEDA employ LLMs to decompose design tasks, generate tool scripts, and orchestrate conventional EDA tools from RTL through physical implementation [8]. Collectively, these efforts exploit three capabilities of LLMs—cross-domain semantic mapping, pattern recognition, and automated reasoning—to bridge high-level design intent with executable artifacts, including HDL code, verification programs, and EDA scripts. Rather than replacing established EDA algorithms, LLMs increasingly serve as intelligent interfaces and assistants that automate repetitive tasks, guide design-space exploration, and enable engineers to focus on architecture, corner cases, and other tasks requiring domain expertise.

\section{SystemVerilog Assertion Generation}
\label{sec:sva-gen}

\begin{table*}[t]
\centering
\caption{Representative related work on LLM-assisted SVA generation across training- and inference-time techniques}
\label{tab:sva-related-work-inference}
\tiny
\setlength{\tabcolsep}{1.5pt}
\renewcommand{\arraystretch}{1.05}
\begin{tabularx}{\textwidth}{P{1.45cm}P{0.85cm}P{1.25cm}YYYY}
\toprule
\textbf{Work} & \textbf{Pre- or Post-RTL} & \textbf{Technique stage} & \textbf{Generation style} & \textbf{Feedback / evaluation} & \textbf{Main contribution} & \textbf{Key limitation} \\
\midrule
nl2sva \cite{nl2sva:hahn:2023} & Post-RTL & Inference & Circuit-aware few-shot prompting with editable sub-translations & JasperGold/model-checker feedback with human review & Early circuit-aware NL-to-SVA translation & Proof-of-concept evaluation; substantial human interaction \\
AssertLLM \cite{assertllm:fang:2024} & Pre-RTL & Inference & Multi-LLM task decomposition with structured text/waveform IRs & JasperGold FPV on I2C; syntax and proof pass rates & Full-specification multimodal generation using intermediate representations & Limited benchmark scope; no verifier-in-the-loop refinement \\
AutoSVA2 \cite{autosva2} & Post-RTL & Inference & GPT-4 RTL prompting with engineered syntax/semantic rules & FPV counterexamples and coverage on CVA6 PTW/TLB & Integrates direct RTL-to-SVA generation with AutoSVA & Requires engineer auditing/refinement and formal-tool access \\
AssertionForge \cite{assertionforge:bai:2025} & Post-RTL & Inference & Spec+RTL knowledge graph with RAG/SSR and GRW-AS retrieval & JasperGold proof plus COI statement/branch/functional/toggle coverage & Structured spec--RTL grounding for broader assertion generation & KG overhead; no verifier-in-the-loop SVA refinement \\
ChiRAAG \cite{chiraag:mali:2024} & Post-RTL & Inference & Structured-spec prompting with iterative simulation-log repair & Simulation/test-case feedback on OpenTitan designs & Early automatic simulator-feedback refinement of generated SVAs & Few test cases; limited completeness/coverage guarantees \\
SANGAM \cite{sangam:gupta:2025} & Post-RTL & Inference & Multimodal agents with Monte Carlo Tree Self-Refine (MCTSr) & JasperGold proof/coverage feedback on I2C and RV-Timer & Search-based self-refinement for multimodal SVA generation & Two-design evaluation; compute/tool-intensive refinement \\
AssertCraft \cite{assertcraft:shahidzadeh:2024} & Post-RTL & Training + Inference & Subtask-focused fine-tuning plus iterative compiler/FPV-guided prompting & Custom compiler and FPV on HDLBits-derived designs & Combines task-specific fine-tuning with automated syntax/semantic repair & Requires curated training data and golden RTL; gains are coupled to refinement loop \\
AssertCoder \cite{assertcoder:tian:2025} & Pre-RTL & Inference & Multimodal extraction plus semantic analyzers and multi-step CoT & Mutation-based model checking and refinement on three designs & Unified text/table/diagram/formula-to-SVA generation & Extraction errors propagate; evaluation limited to three designs \\
Spec2Assertion \cite{spec2assertion:wu:2025} & Pre-RTL & Inference & Progressive regularization with CoT and property decomposition & JasperGold FPV plus RTL-derived assertion-importance score & Explicit pre-RTL generation and assertion-quality metric & Evaluation metric still requires golden RTL \\
AssertGen \cite{assertgen:lyu:2025} & Post-RTL & Inference & CoT objectives with cross-layer spec-to-RTL signal bridging & FPV, COI/proof-core, and mutation coverage on five designs & Generates assertions over internal cross-layer signal chains & Requires RTL/signal artifacts; no closed-loop refinement \\
ChatSVA \cite{chatsva:fu:2026} & Pre-RTL & Training + Inference & Four task-specific agents with AgentBridge data generation, SFT, and RAG & Formal verification on 24 designs; pass rate and coverage & Specialized trained agents for high-coverage end-to-end SVA generation & Heavy data/training pipeline; formal evaluation still depends on RTL oracle \\
SuperSAGA \cite{supersaga:paul:2026} & Post-RTL & Inference & Supervisor--subordinate agents with RAG and verification plan & Iterative formal-tool feedback; OpenTitan coverage evaluation & Agent hierarchy for generation, debug, and refinement & Requires human-reviewed verification plans and formal tools \\
SVAgent \cite{svagent:guo:2025} & Post-RTL & Inference & Security-property decomposition with RTL/FSM analysis & Pyverilog/TrustHUB; syntax and functional accuracy & Security-focused SVA generation with structured reasoning chains & Limited to selected threat models/vulnerability classes \\
Saarthi \cite{saarthi:kumar:2026} & Post-RTL & Inference & Multi-agent vPlan-to-SVA, proof/CEX repair, and coverage closure & Formal proof, counterexamples, coverage, and human escalation & End-to-end AI formal-verification engineer workflow & Human intervention remains for unresolved agent loops \\
ProofLoop \cite{proofloop:mashnoor:2026} & Post-RTL & Inference & ReAct agent with AST-RAG and JasperGold structural queries & Three-round JasperGold proof refinement on FVEval Design2SVA & Adaptive RTL retrieval with solver-in-the-loop generation & Requires retrieval infrastructure and commercial FV tooling \\
KG-Based Agentic FV \cite{kgagenticfv:viswambharan:2026} & Post-RTL & Inference & Spec+RTL+feedback KG with syntax/CEX/coverage refinement loops & Formal diagnostics, CEX, and coverage on seven benchmarks & Persistent traceable grounding across the FV workflow & Complex KG/agent state; design-dependent convergence \\
\midrule
Validatable Generation \cite{finetunefirst:aditi:2023} & Pre-RTL & Training & Fine-tuned CodeT5/PLBART for NL-to-SVA with back-translation validation & Back-translation/cycle-consistency similarity; no RTL formal check & Early supervised NL-to-SVA translation with an interpretable validation loop & Property-level input only; no full-specification extraction or RTL validation \\
VERT \cite{vert:menon:2025} & Post-RTL & Training & LoRA fine-tuning of LLaMA 3.1 and DeepSeek Coder on Verilog/SVA pairs & Syntax plus mutation, formal, and simulation-based functional checks & Large domain-specific SVA dataset showing strong gains from local fine-tuning & RTL-snippet task; limited evidence for full specification-driven generation \\
Hybrid-NL2SVA \cite{hybridnl2sva:xiao:2025} & Pre-RTL & Training + Inference & Qwen2.5-Coder supervised fine-tuning combined with customized HybridRetrieval RAG & Formal evaluation on 40 designs and 229 annotated SVAs & Demonstrates complementary gains from SFT and retrieval for NL-to-SVA & Targets property-level translation rather than property extraction/full-spec generation \\
\bottomrule
\end{tabularx}
\end{table*}

\begin{table*}[t]
\centering
\caption{Representative datasets and benchmarks for LLM-assisted SVA generation}
\label{tab:sva-datasets}
\tiny
\setlength{\tabcolsep}{1.8pt}
\renewcommand{\arraystretch}{1.1}
\begin{tabularx}{\textwidth}{P{1.55cm}P{1.15cm}P{1.35cm}YYYY}
\toprule
\textbf{Dataset / benchmark} & \textbf{Primary use} & \textbf{Scale} & \textbf{Inputs / targets} & \textbf{Construction / source} & \textbf{Evaluation / intended use} & \textbf{Key limitation} \\
\midrule
AssertLLM \cite{assertllm:fang:2024} & Evaluation & 20 designs & Complete specifications + golden RTL; generated SVAs are evaluated against the design & Open-source IP and system-level designs with full natural-language specifications and verified RTL & Syntax, FPV pass/fail, and COI/proof coverage using formal verification & Small design set and assumes bug-free golden RTL; formal success is only a proxy for specification alignment \\
VERT \cite{vert:menon:2025} & Training & 20,000 samples & RTL/code snippets $\rightarrow$ SVA & Synthetic if-else, case, and combined-logic snippets using variables extracted from open-source HDL projects; synchronous and asynchronous cases & Domain-specific fine-tuning of open-source LLMs for SVA generation & Synthetic snippet-level data does not represent complete specifications, modules, or hierarchical design intent \\
AssertionBench \cite{assertionbench:pulavarthi:2025} & Evaluation + training & 100 designs & Verilog design $\rightarrow$ assertion set & Curated OpenCores designs; assertions mined with GoldMine/HARM and formally verified with JasperGold & In-context learning and fine-tuning studies; syntax and FPV proof/counterexample outcomes & RTL-to-SVA setting lacks natural-language specification intent and inherits biases of mined reference assertions \\
FVEval \cite{fveval:kang:2024} & Evaluation & 571 cases & NL/testbench $\rightarrow$ SVA; synthetic NL $\rightarrow$ SVA; RTL $\rightarrow$ SVA & 79 expert NL2SVA cases, 300 synthetic NL2SVA cases, and 192 generated RTL cases (96 pipelines + 96 FSMs) & Syntax and formal functional/equivalence checking across three assertion-generation settings & Human subset is small; two subsets are synthetic and do not model full specification-driven verification \\
Hybrid-NL2SVA \cite{hybridnl2sva:xiao:2025} & Training + evaluation & 4,070 training examples; 40 designs / 229 SVAs for evaluation & Natural-language property/explanation $\rightarrow$ SVA & Synthetic fine-tuning corpus with prompt-guided SVA construction explanations; separate formally verified annotated evaluation set & Fine-tuning/RAG evaluation using syntax and functional/formal matching & Focuses on property-level NL2SVA translation rather than extracting properties from complete specifications or RTL \\
CVDP-cid14 \cite{cvdp:pinckney:2025} & Evaluation & 98 assertion tasks & Verification prompt/context $\rightarrow$ SVA & Human-authored tasks within the 783-problem CVDP suite; 68 non-agentic and 30 agentic assertion-generation problems & Executable test-harness/simulation-based pass evaluation in realistic single- and multi-step settings & Assertion generation is only one CVDP category and simulation/harness success does not establish formal or semantic correctness \\
AssertLLM2 \cite{assertllm2:wu:2026} & Evaluation & 83 designs, 13 categories & Structured specification + golden/buggy RTL $\rightarrow$ SVA & Real-world open-source designs (filtered for $>200$ RTL lines), normalized specifications, dependency-complete RTL, and systematic mutations & Syntax, formal provability, COI/proof/formal coverage, and mutation-based bug detection in bug-prevention and bug-hunting settings & Still relatively small and untested among the broader LLM for SVA generation community \\
\bottomrule
\end{tabularx}
\end{table*}

This section first overviews SVA generation and the motivations of using LLMs.
Then, it survyes recent LLM-based SVA generation methods at inference time and training time as summarzed in Table~\ref{tab:sva-related-work-inference}.
Next, it reviews evaluation methods and bechmarks for SVA generation methods as summarized in Table~\ref{tab:sva-datasets}.
Finally, remaining challenges and future directions are highlighted. 

\subsection{Overview and Motivation}
\label{sec:sva-overview}


Assertion-Based Verification (ABV) expresses expected hardware behavior as executable properties that can be evaluated through simulation or formally proven against an implementation. SystemVerilog Assertions (SVAs) are the dominant assertion language, used in approximately 80\% of projects employing ABV~\cite{vstudy:foster:2024}. SVAs capture functional and temporal requirements, such as protocol rules, request--response relationships, ordering constraints, and reset behavior, thereby providing an explicit representation of design intent. They improve observability and debugging and enable formal exploration of behaviors that may be difficult to exercise through simulation.

Despite these advantages, developing high-quality SVAs remains knowledge-intensive and time-consuming. Verification engineers must interpret requirements distributed across natural-language specifications, tables, diagrams, and timing descriptions; identify the behaviors that require verification; map abstract concepts to implementation signals; and encode them using precise temporal semantics. Ambiguous or incomplete specifications further widen this semantic gap between high-level design intent and executable assertions.

Accordingly, automated SVA generation must address more than syntactic correctness. 
A useful assertion must be \emph{syntactically valid}, \emph{semantically consistent with the intended requirement}, and \emph{correctly aligned with implementation signals and cycle-level timing}. 
These requirements are distinct: an assertion may compile successfully yet express the wrong behavior, while an assertion formally proven against RTL may still misrepresent the specification. Signal and timing alignment is particularly challenging because specifications often describe behavior abstractly, whereas SVAs must identify concrete signals, sampling events, delays, and temporal relationships. Beyond individual assertions, practical generation should produce an assertion set that provides adequate specification coverage while avoiding vacuous, redundant, or inconsistent properties.

\subsubsection{Automated SVA Generation Before LLMs}

Before LLMs, automated assertion generation primarily relied on templates, rules, static analysis, and mining of simulation traces. GoldMine~\cite{goldmine:vasudevan:2010} mines candidate invariants from simulation traces, maps them to predefined assertion templates, and applies formal verification to eliminate invalid candidates. A-Team~\cite{ateam:danese:2017} combines templates with data mining and coverage-guided heuristics, while HARM~\cite{harm:germiniani:2022} uses user-provided propositions, templates, and metrics to generate temporal properties.

These approaches provide predictable and structured generation, but their expressiveness is constrained by predefined templates, propositions, or observed execution behavior. Consequently, they can struggle with previously unseen property structures, complex temporal relationships, and requirements expressed only at the specification level. Trace-mining methods may also infer properties that happen to hold for the observed implementation rather than properties representing intended behavior. More fundamentally, conventional approaches have limited capability to interpret unstructured specifications and determine \emph{what properties should be verified}, leaving a semantic gap between design requirements and generated assertions.

\subsubsection{Motivation for LLM-Based SVA Generation}

LLMs offer a complementary capability by interpreting natural-language requirements, generating code, and reasoning across heterogeneous design information. They can potentially extract behavioral relationships from specifications, translate them into temporal properties, and, when RTL is available, map specification concepts to concrete signals and implementation structures. Retrieval and multimodal techniques can further incorporate information from large documents, tables, diagrams, and waveforms. These capabilities make LLMs particularly attractive for bridging the specification-to-assertion semantic gap.

However, LLM generation is probabilistic and provides no correctness guarantee. Generated assertions may contain syntax errors, incorrect temporal relationships, erroneous signal mappings, or properties that appear plausible but do not reflect design intent. Recent approaches therefore increasingly combine LLM generation with structured design representations, retrieval, simulation, formal verification, coverage analysis, and iterative refinement. In such workflows, LLMs provide semantic interpretation and candidate generation, while deterministic verification tools provide feedback for validation and refinement.

A key distinction is whether assertions are generated before or after RTL becomes available. \emph{Pre-RTL} approaches derive properties primarily from specifications and therefore preserve greater independence from implementation errors, but face difficulties in mapping abstract requirements to concrete signals and cycle-level timing. \emph{Post-RTL} approaches use implementation information to improve signal and timing grounding and produce more readily executable assertions, but risk generating properties that reproduce erroneous RTL behavior rather than independently checking the specification. Thus, effective LLM-based SVA generation must balance \emph{specification fidelity} with \emph{implementation grounding}, using RTL to resolve implementation details without allowing it to override the verification intent established by the specification.


%
\subsection{Inference-Time Methods}
\label{sec:sva-inference-time}

Inference-time approaches generate SVAs using pretrained LLMs without modifying their model parameters. Compared with fine-tuning-based methods, they can readily incorporate new specifications, RTL designs, retrieved knowledge, and verification-tool feedback. 
Existing methods have evolved from direct natural-language translation toward structured, multimodal, tool-grounded, and agentic workflows that address different stages of the specification-to-SVA transformation.

\subsubsection{Structured Decomposition and Design Grounding}

NL2SVA~\cite{nl2sva:hahn:2023} is among the earliest LLM-based approaches and translates manually provided natural-language property descriptions into an intermediate representation containing signal information, partial translations, and candidate SVAs. Model checking is subsequently used to validate the assertions, with human intervention when refinement is required. Although this workflow demonstrates the feasibility of combining LLM generation with formal verification, it assumes that relevant requirements have already been extracted from the specification and therefore does not automate the more difficult task of identifying verification properties from complete design documents.

AssertLLM~\cite{assertllm:fang:2024} extends this direction toward end-to-end generation from complete specifications. Rather than directly prompting an LLM with the entire document, it decomposes the task into specification analysis, signal mapping, and assertion generation, thereby producing structured intermediate descriptions before generating SVAs. It also incorporates assertion-related knowledge through retrieval and processes textual and waveform information. This decomposition reduces the reasoning burden of a single inference step and makes intermediate results more inspectable. However, because AssertLLM derives its design knowledge primarily from the specification, implementation details absent from the specification can remain unresolved.

AssertionForge~\cite{assertionforge:bai:2025} addresses this limitation by jointly grounding generation in the specification and RTL. It constructs a hardware-specific knowledge graph (KG) from specification concepts and fuses it with RTL entities and relationships, followed by multi-resolution context extraction for SVA generation. Compared with the primarily specification-centered decomposition of AssertLLM, this representation explicitly connects high-level requirements with implementation-level signals and dependencies, improving the context available for assertion generation. The tradeoff is increased preprocessing complexity and greater dependence on the RTL: errors in KG construction or an incorrect implementation can propagate into generated properties, potentially weakening their independence as specification-level checks.

AutoSVA2~\cite{autosva2} similarly introduces intermediate rules to constrain SVA generation. Collectively, these approaches show that intermediate representations---including structured descriptions, rules, and knowledge graphs---are useful for decomposing the large semantic jump from heterogeneous specifications to temporal logic. However, additional representations also create new points at which information can be omitted or incorrectly transformed.

\subsubsection{Multimodal Processing and Explicit Reasoning}

Hardware specifications frequently encode critical behavior in tables, timing diagrams, waveforms, equations, and figures rather than text alone. AssertLLM~\cite{assertllm:fang:2024} incorporates waveform information, while AssertCoder~\cite{assertcoder:tian:2025} more explicitly decomposes multimodal specification processing by converting diagrams, waveforms, tables, and formulas into structured representations such as pseudocode, formal expressions, \LaTeX, or JSON before assertion generation. SANGAM~\cite{sangam:gupta:2025} likewise employs specialized Spec Analyzer, Signal Mapper, and Waveform Analyzer components to extract complementary information from specifications. These approaches reduce dependence on textual descriptions and can recover timing or structural information unavailable in prose. Their effectiveness, however, depends on the accuracy of multimodal interpretation; an incorrectly interpreted waveform or table can introduce semantic errors before SVA generation begins.

Explicit reasoning provides another mechanism for bridging specifications and temporal properties. Spec2Assertion~\cite{spec2assertion:wu:2025} applies Chain-of-Thought (CoT) prompting and progressive regularization in the pre-RTL setting, encouraging the model to reason about antecedents, consequents, temporal relationships, and cycle dependencies before constructing an SVA. AssertCoder~\cite{assertcoder:tian:2025} extends multi-step reasoning to structured multimodal inputs, whereas AssertGen~\cite{assertgen:lyu:2025} first extracts verification objectives from the specification and subsequently maps them to RTL signals. The latter approaches therefore combine explicit reasoning with implementation grounding, addressing a major limitation of purely pre-RTL reasoning. Nevertheless, a plausible reasoning trace is not evidence of semantic correctness: incorrect timing assumptions or incomplete interpretation of the requirement can still produce well-formed but incorrect assertions.

\subsubsection{Retrieval and Verification-Guided Refinement}

Retrieval-Augmented Generation (RAG) provides external knowledge without modifying the LLM. AssertLLM~\cite{assertllm:fang:2024}, for example, retrieves SVA-related verification knowledge to improve generation, whereas AssertionForge~\cite{assertionforge:bai:2025} uses graph-based retrieval over design-specific relationships. The distinction is important: general verification knowledge primarily helps the model understand \emph{how to express} assertions, while design-specific retrieval helps determine \emph{what behavior should be expressed}. The latter is therefore more directly relevant to specification-to-RTL semantic alignment.

Most recent frameworks further couple generation with iterative feedback. NL2SVA~\cite{nl2sva:hahn:2023} introduced model-checking feedback with human-guided refinement, while general self-refinement methods such as Self-Refine~\cite{selfrefine:madaan:2023} demonstrated that LLMs can critique and revise their own outputs. ChIRAAG~\cite{chiraag:mali:2024} applies verification-tool feedback to SVA correction by interpreting errors and regenerating failed assertions. Such feedback is particularly effective for syntax and tool-detectable errors because compilers and formal engines provide objective diagnostics. Its ability to correct semantic errors is more limited: a property that successfully compiles or is proven against the RTL may still encode the wrong requirement.

SANGAM~\cite{sangam:gupta:2025} extends linear self-refinement into candidate search through Monte Carlo Tree Self-Refine (MCTSr). After multimodal specification processing, it explores multiple reasoning trajectories for individual signals and uses the resulting traces to generate assertions. Compared with one-shot or linear refinement, this approach reduces dependence on a single stochastic generation and can explore alternatives involving different signal dependencies or temporal relationships. However, its effectiveness depends critically on the reward or selection criterion; optimizing primarily for formal-tool success can favor assertions that are provable but weak, vacuous, or semantically misaligned. SANGAM therefore represents a shift from simply \emph{repairing} a generated assertion toward \emph{searching} the space of candidate interpretations.

AssertCoder~\cite{assertcoder:tian:2025} and AssertGen~\cite{assertgen:lyu:2025} further incorporate mutation testing to evaluate whether generated assertions detect injected RTL faults. This provides stronger evidence of verification usefulness than syntax or proof success alone because it evaluates fault-detection capability. Nevertheless, mutation score remains dependent on the representativeness of the injected faults and does not by itself establish that an assertion faithfully represents the specification.

\subsubsection{Agentic SVA Generation}

The progression from task decomposition and iterative refinement naturally leads to agentic workflows, in which specialized components coordinate specification analysis, retrieval, assertion generation, tool execution, and repair. AssertLLM~\cite{assertllm:fang:2024} and SANGAM~\cite{sangam:gupta:2025} already exhibit elements of role-based decomposition, although their primary contributions are structured generation and search rather than autonomous tool orchestration.

More explicit agentic approaches extend this decomposition. ChatSVA~\cite{chatsva:fu:2026} formulates SVA generation as an end-to-end multi-agent workflow with specialized stages for specification processing, functional-checkpoint extraction, and assertion generation, while its AgentBridge infrastructure supports construction of task-specific data. SuperSAGA~\cite{supersaga:paul:2026} uses a supervisor--subordinate architecture combining RAG, human-reviewed verification plans, formal-tool feedback, and iterative assertion refinement. SVAgent~\cite{svagent:guo:2025} applies agentic decomposition specifically to hardware-security verification by breaking complex security requirements into finer-grained verification tasks before SVA generation.

This direction also overlaps increasingly with broader agentic formal-verification systems. Saarthi~\cite{saarthi:kumar:2026}, ProofLoop~\cite{proofloop:mashnoor:2026}, and KG-based agentic formal-verification frameworks~\cite{kgagenticfv:viswambharan:2026} coordinate assertion generation with mechanisms such as verification rulebooks, knowledge-graph retrieval, formal-tool queries, counterexample-guided refinement, and coverage feedback. These systems expand the role of the LLM from an assertion generator to an orchestrator of the verification process. 
However, greater autonomy also increases pipeline complexity, inference and tool-execution cost, and opportunities for error propagation. 
Consequently, agentic decomposition is most useful when intermediate artifacts such as extracted requirements, signal mappings, verification plans, and candidate assertions can themselves be independently validated rather than trusting only the final output.

\subsubsection{Summary and Remaining Limitations}

Overall, inference-time SVA generation has evolved along several complementary dimensions. Early methods such as NL2SVA focus on translating explicitly provided requirements, whereas AssertLLM expands generation to complete specifications through task decomposition. AssertionForge and AssertGen strengthen specification-to-RTL grounding; AssertCoder and SANGAM incorporate multimodal information; Spec2Assertion emphasizes explicit reasoning; SANGAM further introduces search over alternative reasoning trajectories; and recent agentic frameworks coordinate these capabilities with external verification tools. Thus, the field is moving from direct \emph{NL-to-SVA translation} toward \emph{specification understanding, design grounding, candidate generation, verification, and iterative refinement}.

Despite this progression, three limitations remain common. First, stronger RTL grounding improves signal mapping and executability but can compromise independence from implementation errors, whereas specification-only methods preserve verification intent but struggle with implementation-level details. Second, tool feedback provides reliable evidence for syntax, satisfiability, and behavior relative to the RTL, but does not by itself establish semantic agreement with the specification. Third, increasingly elaborate reasoning, search, and multi-agent pipelines improve robustness at the expense of latency, tool calls, computational cost, and additional opportunities for intermediate errors. Future inference-time methods therefore need to combine specification fidelity and implementation grounding while independently validating intermediate representations and evaluating generated assertion sets using complementary measures of semantic correctness, coverage, non-vacuity, and realistic fault-detection capability.
%
\subsection{Training-Time Methods}
\label{sec:sva-training-time}

Training-time approaches adapt an LLM to SVA generation by updating its parameters using hardware-verification-specific data. Unlike inference-time methods, which rely on prompting, retrieval, structured context, search, or tool feedback without modifying the underlying model, fine-tuning attempts to internalize SVA syntax, temporal semantics, and recurring hardware-property patterns. This can improve generation consistency and enable smaller open-source models to perform specialized verification tasks. However, because SVA generation involves multiple stages---including requirement extraction, signal mapping, temporal reasoning, and assertion construction---most existing training-time methods focus on particular subtasks rather than learning the complete specification-to-SVA process.

\subsubsection{Property-Level Translation and Semantic Adaptation}

Aditi and Hsiao~\cite{finetunefirst:aditi:2023} provide an early demonstration of fine-tuning for SVA generation by adapting CodeT5 and PLBart to translate natural-language property descriptions into assertions. The generated SVA is additionally translated back into natural language and compared with the original description, providing an interpretable consistency check between the input requirement and generated property. This work demonstrates that task-specific training can teach models the correspondence between natural-language temporal behavior and SVA constructs. However, it assumes that the desired property has already been identified and expressed as a concise natural-language statement. It therefore addresses \emph{property translation} rather than the broader problem of extracting verification requirements from complete specifications. Moreover, back-translation measures linguistic consistency rather than behavioral correctness; an assertion can appear consistent in natural language while still containing incorrect temporal semantics or failing to represent the RTL behavior correctly.

Hybrid-NL2SVA follows a similar property-translation formulation but combines domain-specific fine-tuning with inference-time augmentation. It fine-tunes Qwen2.5-Coder-7B-Instruct using a synthetic dataset constructed from hardware-design textbooks and Verilog collected from 67 designs, containing 4,070 SVAs paired with one-sentence natural-language descriptions generated using OpenAI o4-mini. The resulting model shows substantial improvement after fine-tuning, while retrieval-based augmentation further improves generation. Compared with the earlier CodeT5/PLBart approach, Hybrid-NL2SVA benefits from a stronger pretrained code model, a larger hardware-specific dataset, and a hybrid training/inference strategy. Nevertheless, both approaches share the same fundamental abstraction: they assume that a relevant property has already been extracted and expressed in natural language. Their results therefore demonstrate improved \emph{NL-to-SVA translation}, but provide limited evidence for full specification-driven assertion generation. In addition, synthetic natural-language labels and automatically collected assertions make performance dependent on the semantic quality of the constructed training pairs.

\subsubsection{Specification-Conditioned and Hybrid Fine-Tuning}

AssertCraft~\cite{assertcraft:shahidzadeh:2024} moves closer to specification-driven generation by fine-tuning an LLM using assertion-related examples extracted and filtered from open-source repositories together with synthetic data. Rather than relying exclusively on the fine-tuned model, AssertCraft combines task-specific adaptation with an inference-time iterative refinement loop in which candidate assertions can be repaired using a non-fine-tuned LLM. This hybrid architecture is significant because it treats fine-tuning and inference-time reasoning as complementary: fine-tuning provides domain-specific generation capability, while iterative refinement addresses errors that remain after generation.

This design also illustrates an important difficulty in evaluating training-time approaches. Improvements in AssertCraft arise from several interacting components, including dataset construction, task decomposition, the fine-tuned generator, and iterative refinement. Consequently, end-to-end improvements cannot necessarily be attributed to fine-tuning alone without controlled ablation. More generally, fine-tuning can improve the model's prior knowledge of SVA patterns but does not eliminate the need for specification grounding or external verification. A model trained on high-quality assertion examples may still generate a property that is inappropriate for the particular design being verified.

\subsubsection{Domain-Specific Model Adaptation}

VERT~\cite{vert:menon:2025} further emphasizes the value of domain-specific training data. It constructs a large SVA dataset from open-source hardware designs and uses it to fine-tune DeepSeek Coder and LLaMA~3.1 models, which are subsequently compared with GPT-4o across multiple hardware designs. The results show that smaller open-source models adapted to hardware-verification data can outperform a substantially larger general-purpose closed-source model on the evaluated SVA-generation tasks. This provides important evidence that, for specialized verification tasks, domain adaptation can compensate for model scale. Fine-tuned open-source models also enable local deployment, which is attractive when proprietary RTL and specifications cannot be exposed to external model providers.

VERT, however, also exposes a recurring limitation of current training-time methods: the granularity and provenance of training data strongly determine what the model learns. Assertions mined from open-source repositories provide realistic SVA syntax and recurring temporal patterns, but snippet-level examples contain limited information about the complete specification and design context from which the assertions originated. Consequently, strong performance on local assertion-generation tasks does not necessarily imply that the model can identify appropriate verification objectives from a full specification, reason across module hierarchies, or construct a comprehensive assertion set for an unseen IP.

\subsubsection{Cross-Method Analysis and Limitations}

Taken together, these studies demonstrate a progression from fine-tuning for direct NL-to-SVA translation toward domain-adapted models embedded within larger verification pipelines. Aditi and Hsiao~\cite{finetunefirst:aditi:2023} establish the feasibility of learning mappings between natural-language properties and SVA; Hybrid-NL2SVA scales this formulation using a modern code LLM, synthetic data, and retrieval; AssertCraft~\cite{assertcraft:shahidzadeh:2024} combines fine-tuning with iterative inference-time refinement; and VERT~\cite{vert:menon:2025} demonstrates that sufficiently specialized training can make smaller open-source models competitive with or superior to larger general-purpose models. Thus, training-time and inference-time techniques are increasingly complementary rather than competing alternatives: fine-tuning internalizes recurring domain knowledge, while retrieval, structured context, RTL grounding, and verification-tool feedback provide design-specific information that cannot reasonably be encoded in model parameters.

Nevertheless, current training-time methods remain primarily focused on learning \emph{how to express properties as SVAs}, rather than determining \emph{which properties are required to adequately verify a design}. Most training datasets consist of assertion snippets or pairs of natural-language descriptions and SVAs, with relatively limited representation of complete specifications, RTL hierarchy, verification intent, and relationships among multiple properties. Dataset construction also introduces risks from noisy repository assertions, automatically generated descriptions, duplicated examples, and possible overlap between training and evaluation designs. These issues make data quality, provenance, and contamination control particularly important when evaluating fine-tuned models.

Future training-time methods therefore require datasets that preserve richer relationships among \emph{specifications, RTL, verification objectives, assertions, and verification outcomes}. Training objectives could similarly move beyond next-token prediction on individual assertions toward learning specification-to-property alignment, signal grounding, temporal reasoning, and assertion-set construction. Importantly, fine-tuning should not replace deterministic verification: compilation, simulation, formal analysis, vacuity checking, and mutation testing remain necessary for independently evaluating generated properties. A promising direction is therefore a hybrid architecture in which domain-specific training provides efficient hardware-verification knowledge, while inference-time retrieval and reasoning supply design-specific context and formal tools provide objective feedback on correctness and verification effectiveness.
%
\subsection{Evaluation Methods and Benchmarks}
\label{sec:sva-evaluation}

The inference- and training-time methods reviewed above employ different mechanisms to improve SVA generation, ranging from structured prompting, retrieval, reasoning, and formal-tool feedback to domain-specific fine-tuning. Evaluating these methods, however, requires determining not only whether an LLM can produce executable SVA code, but also whether the generated properties faithfully capture verification intent and effectively detect incorrect behavior. Existing evaluations can therefore be viewed as providing increasingly strong but complementary forms of evidence: \emph{syntactic validity}, \emph{formal consistency}, \emph{coverage}, \emph{fault-detection capability}, and ultimately \emph{specification alignment}.

\subsubsection{Evaluation Methodologies}

\textbf{Syntactic validity} is the most basic and widely reported metric. It measures whether generated assertions can be parsed, compiled, and accepted by SystemVerilog verification tools. This metric is inexpensive, deterministic, and particularly useful for measuring whether prompting or fine-tuning improves knowledge of SVA syntax. However, compilation establishes only that an assertion is executable; a syntactically correct property may still check the wrong behavior, use inappropriate timing, or encode an incorrect requirement.

\textbf{Formal property verification (FPV)} provides stronger behavioral evidence by checking generated assertions against an RTL implementation using formal verification tools such as JasperGold or SymbiYosys. If a trusted golden RTL implementation is assumed to correctly realize the specification, successful proof provides evidence that the generated assertion is consistent with the expected design behavior, while counterexamples can identify incorrect properties. FPV is therefore widely used in benchmarks such as AssertLLM~\cite{assertllm:fang:2024}, VERT~\cite{vert:menon:2025}, AssertionBench~\cite{assertionbench:pulavarthi:2025}, and FVEval~\cite{fveval:kang:2024}. However, FPV is not a direct semantic oracle. A weak, vacuous, or incorrectly interpreted assertion may still be provable, and results depend on the RTL, assumptions, constraints, and formal environment. Proofs may also time out or remain inconclusive for complex designs. Thus, FPV primarily establishes \emph{consistency with the analyzed implementation}, not necessarily correctness with respect to the original specification.

\textbf{Coverage analysis} complements FPV by measuring how generated properties relate to the implementation. Existing work reports metrics such as cone of influence (COI), proof coverage, formal coverage, and, depending on the environment, statement, branch, or toggle coverage. AssertLLM~\cite{assertllm:fang:2024} introduced COI-based analysis alongside FPV, while later benchmarks such as VERT~\cite{vert:menon:2025} and AssertLLM2~\cite{assertllm2:wu:2026} employ broader coverage measurements. Coverage can expose trivial assertions that constrain or observe little of the design and can assess the collective reach of an assertion set. Nevertheless, high coverage does not imply correct verification intent: a property may involve many RTL signals while still encode behavior inconsistent with the specification.

\textbf{Mutation testing} evaluates a different dimension---whether assertions can detect erroneous implementations. Mutations are introduced into an otherwise correct RTL design, and generated assertions are evaluated according to whether they fail on the faulty versions. AssertCoder~\cite{assertcoder:tian:2025}, VERT~\cite{vert:menon:2025}, and AssertLLM2~\cite{assertllm2:wu:2026} use mutation-based evaluation to complement syntax, proof, and coverage metrics. Compared with proving assertions only against golden RTL, mutation testing provides stronger evidence of practical fault-detection capability and assertion strength. However, its validity depends strongly on the realism, diversity, and relevance of the injected faults. Easily detected or artificial mutations may inflate performance, while generating representative semantic bugs at scale remains difficult. Mutation score should therefore be interpreted as a measure of \emph{fault-detection effectiveness}, rather than direct proof of specification correctness.

These metrics evaluate complementary properties rather than interchangeable notions of correctness. Syntax checking establishes \emph{executability}; FPV establishes \emph{consistency with RTL under the verification environment}; coverage measures \emph{interaction with or reach into the design}; and mutation testing measures \emph{ability to detect selected faults}. The strongest objective, \textbf{specification alignment}, asks whether an assertion accurately captures the intended requirement. This remains the most difficult dimension to evaluate automatically because neither the RTL nor generated mutations provide an independent semantic oracle when the objective is to verify the implementation against a natural-language specification. Consequently, no single existing metric is sufficient, motivating evaluation frameworks that combine multiple complementary measures.

\subsubsection{Benchmarks}

Existing SVA benchmarks differ substantially in task formulation, design scale, input context, and evaluation methodology, reflecting the evolution from property translation toward specification- and design-level assertion generation.

AssertLLM~\cite{assertllm:fang:2024} introduced one of the earliest standardized benchmarks for specification-driven SVA generation. It contains 20 samples drawn from open-source IP and system-level designs, pairing specifications with corresponding RTL. This pairing is important because it enables evaluation of both specification-conditioned generation and generated assertions against executable implementations. Its evaluation combines syntax checking, COI analysis, and FPV, providing stronger evidence than syntax-only evaluation. However, the small number of designs limits behavioral and structural diversity and makes broad conclusions about generalization difficult.

VERT~\cite{vert:menon:2025} takes a substantially different approach. Its dataset contains approximately 20,000 snippets extracted from open-source hardware repositories, including BOOM, Rocket Chip, XiangShan, and other RTL projects. Rather than emphasizing complete specification-to-design tasks, VERT targets recurring SVA-generation patterns and errors, including temporal and clock-cycle interpretation. Its scale makes it particularly suitable for training and evaluating domain-adapted models, and its combination of syntax, COI, FPV, and mutation testing provides multiple measures of assertion quality. The tradeoff is granularity: snippet-level tasks provide limited evidence that a model can identify verification objectives from a complete specification or reason across large hierarchical designs. Thus, VERT is especially valuable for evaluating the \emph{training-time domain adaptation} discussed in the previous subsection, whereas specification-level benchmarks better exercise the inference-time reasoning and grounding pipelines reviewed earlier.

AssertionBench~\cite{assertionbench:pulavarthi:2025} occupies an intermediate position. It contains 100 samples derived from OpenCores designs, with formally proven reference assertions, English descriptions, and corresponding Verilog designs. This construction provides controlled NL-to-SVA tasks while retaining executable RTL for formal validation, making it useful for evaluating in-context learning and property translation. However, its emphasis on syntax and FPV provides limited measurement of assertion-set coverage, fault-detection capability, or fidelity to complete design specifications.

FVEval~\cite{fveval:kang:2024} provides broader task diversity through three complementary sub-benchmarks. NL2SVA-Human contains 79 expert-written assertions from real-world verification environments paired with natural-language descriptions; NL2SVA-Machine contains 300 machine-generated natural-language/SVA pairs; and Design2SVA contains 192 synthetic RTL instances based on arithmetic pipelines and finite-state machines. This decomposition is useful because it distinguishes human-authored property translation, scalable synthetic translation, and direct RTL-to-SVA generation rather than collapsing them into a single task. It therefore enables analysis of how performance changes with the source and form of the input. However, its primary reliance on syntax checking and FPV again leaves specification coverage, realistic fault detection, and full-document semantic alignment largely unmeasured.

CVDP~\cite{cvdp:pinckney:2025} differs from these assertion-specific benchmarks by evaluating SVA generation as one component of a broader suite of LLM-based EDA tasks. Its assertion subset contains 98 tasks with prompts, relevant context, design files, and test harnesses, using checklist-based evaluation and syntax checking. This task-oriented formulation is valuable for assessing SVA generation within a broader hardware-engineering context rather than an isolated translation setting. However, its evaluation provides less evidence about formal correctness, non-vacuity, coverage, and fault-detection strength than assertion-specific benchmarks.

AssertLLM2~\cite{assertllm2:wu:2026} moves toward a more comprehensive design-level evaluation. It contains 83 samples collected from open-source Verilog/SystemVerilog repositories together with corresponding specifications, while filtering out designs below 200 lines of RTL to increase task complexity and normalizing specification structure. More importantly, its evaluation combines syntax checking, FPV, COI, proof coverage, formal coverage, and mutation testing. Compared with earlier benchmarks that emphasize one or two dimensions, AssertLLM2 therefore evaluates both whether assertions are executable and provable and whether they exercise meaningful portions of the design and detect injected faults. Its remaining limitation is shared by the broader literature: these tool-based metrics remain proxies for whether generated assertions faithfully and comprehensively represent the original specification.

\subsubsection{Cross-Benchmark Analysis and Open Challenges}

Taken together, these benchmarks reveal a progression in both \emph{task complexity} and \emph{evaluation strength}. Property-level benchmarks such as AssertionBench and portions of FVEval provide controlled evaluation of NL-to-SVA translation, while VERT emphasizes scale and is particularly suited to training-time domain adaptation. AssertLLM and AssertLLM2 move toward complete specification- and design-conditioned generation, better exercising the reasoning, retrieval, signal-grounding, and refinement capabilities of inference-time systems. CVDP instead evaluates assertion generation within a broader EDA task environment. Similarly, evaluation has progressed from syntax checking toward FPV, coverage, and mutation testing, with recent benchmarks increasingly combining these metrics.

However, direct comparison across reported results remains difficult because benchmarks differ in design source, scale, input information, property granularity, formal environment, and evaluation criteria. Training-time evaluations introduce the additional concern of contamination or repository overlap, particularly when both training and evaluation examples originate from public RTL repositories. Inference-time methods face a related concern when specifications, RTL, or reference assertions may have appeared in model pretraining data. Evaluation on held-out designs and, where possible, previously unpublished or carefully contamination-controlled benchmarks is therefore important for establishing generalization.

More fundamentally, existing benchmarks still provide limited measurement of \emph{assertion-set quality}. Evaluating assertions independently does not determine whether the generated set adequately covers the specification, contains redundant or contradictory properties, or omits critical requirements. Future benchmarks should therefore preserve explicit relationships among specifications, RTL, reference verification objectives, assertions, and realistic faults, enabling separate measurement of syntactic validity, semantic fidelity, specification coverage, non-vacuity, consistency, behavioral diversity, and fault-detection capability. For iterative and agentic methods, evaluation should additionally report inference cost, verification-tool runtime, number of refinement iterations or tool calls, robustness across repeated runs, and human intervention. Such multidimensional evaluation is necessary to compare increasingly sophisticated inference- and training-time approaches and to determine whether improvements in SVA generation translate into more effective hardware verification.
\subsection{Challenges and Future Directions}
\label{sec:sva-challenges}

Despite substantial progress from direct NL-to-SVA translation toward structured, tool-guided, fine-tuned, and agentic approaches, several challenges remain before LLM-generated SVAs can be reliably used in practical verification.

\textbf{Semantic alignment and evaluation.}
A fundamental challenge is determining whether generated assertions faithfully represent the specification. Existing methods commonly rely on syntax checking, formal property verification (FPV), coverage, and mutation testing. These provide complementary evidence of executability, consistency with RTL, design interaction, and fault-detection capability, respectively, but none directly establishes specification fidelity. SpecAlign~\cite{specalign:2026} represents an initial step toward explicitly evaluating the semantic alignment between generated SVAs and their corresponding specifications. Nevertheless, specification-level evaluation remains challenging, particularly when golden RTL is used for both generation and validation, since an incorrect property derived from faulty RTL may still be formally proven against that RTL. Future evaluation should therefore combine complementary metrics with independent specification-level assessment. Standardized benchmarks should also include diverse unseen designs, realistic specifications and faults, and contamination controls to enable meaningful comparison across methods.

\textbf{From individual assertions to assertion sets.}
Existing evaluations primarily assess individual properties, whereas practical verification depends on whether the complete assertion set adequately captures design intent. Generating many simple and correct assertions may produce favorable aggregate results while missing complex behaviors or interactions. Future evaluation should therefore consider specification coverage, non-vacuity, redundancy, consistency, behavioral diversity, and fault-detection effectiveness at the assertion-set level. Explicit traceability between specification requirements and generated properties would further reveal missing or inadequately represented requirements.

\textbf{Generalization, scalability, and cost.}
Training-time approaches show the benefits of domain adaptation, while inference-time methods improve design-specific reasoning through retrieval, structured representations, and tool feedback. However, both remain sensitive to data quality and are largely evaluated on relatively small designs or localized properties. Scaling to complex SoCs requires reasoning across hierarchical RTL and specifications while selectively retrieving relevant information across modules and abstraction levels. Moreover, multimodal processing, candidate search, agentic workflows, and repeated formal-tool calls increase latency and computational cost. Future evaluations should therefore report inference and verification runtime, tool calls, refinement iterations, and human intervention, together with controlled ablations that isolate the contributions of individual techniques.

\textbf{Toward complete verification property sets.}
Another promising direction is joint generation of \emph{assumptions, assertions, and cover properties}. Assumptions characterize valid environmental behavior, assertions specify required design behavior, and cover properties establish reachability of important scenarios. Generating and validating them together can reduce spurious failures, expose overconstraint and vacuity, and provide a more complete representation of verification intent than assertions alone.

Overall, future research should shift from optimizing the generation of individually valid assertions toward constructing \emph{specification-aligned and independently validated verification property sets}. LLMs are well suited for interpreting requirements, connecting information across heterogeneous artifacts, and proposing candidate properties, while deterministic simulation and formal tools provide essential validation and feedback. Combining these complementary capabilities---while maintaining traceability to the specification, controlling cost, and preserving independence from implementation errors---is a promising path toward reducing manual assertion-development effort and enabling reliable, review-centered assertion-based verification.


\section{Testbench Generation}
\label{sec:tb-gen}

This section surveys recent work on automatic testbench and stimulus generation for hardware verification, with emphasis on large language model (LLM) based methods.
It situates LLM approaches relative to pre-LLM template and grammar-based tools, then reviews prompt-only generation, EDA-feedback loops, coverage-directed stimulus generation, self-validating testbench construction, UVM/subsystem-level generation, training-time alignment, and benchmark design.
The main conclusion is that current LLMs are most useful when embedded in verification-aware workflows that constrain generation with scenarios, templates, simulators, coverage reports, and functional self-checks, rather than when used as unconstrained SystemVerilog authors.

\subsection{Motivation and Scope}

Simulation-based functional verification remains one of the most common pre-silicon verification techniques.
A useful testbench is not merely code that compiles: it instantiates the design under test (DUT), drives meaningful stimuli, observes outputs, compares behavior against an oracle or reference model, and records structural or functional coverage.
In industrial SystemVerilog flows, this often expands into Universal Verification Methodology (UVM) components such as sequence items, sequencers, drivers, monitors, agents, scoreboards, coverage collectors, and tests \cite{ieee1800_2023,uvmguide2015}.
Python-based harnesses such as cocotb are also relevant because many LLM workflows exploit Python's stronger representation in pretraining corpora to implement checkers, reference models, or orchestration code \cite{cocotbdocs2026}.

The literature on LLM-assisted testbench generation separates into three related but distinct tasks.
The first is \emph{stimulus generation}: producing input vectors, transactions, or sequences that improve code or functional coverage.
The second is \emph{testbench construction}: producing executable drivers, clocks, resets, monitors, and checkers.
The third is \emph{verification-environment synthesis}: generating reusable UVM or subsystem-level verification infrastructure.
These tasks differ in difficulty.
Stimulus generation can be judged by coverage and simulator output, while full testbench generation also requires correct timing, reset handling, protocol compliance, and a faithful oracle.
This distinction explains why many recent systems use the LLM as a planner or repair agent, while deterministic code templates, DSLs, or simulators enforce the final executable structure \cite{autobench2024,correctbench2025,haven2026,uvmarvel2026}.

\begin{figure*}[t]
\centering
\begin{tikzpicture}[node distance=0.65cm and 0.8cm]
\node[diagramhead, minimum width=5.1cm] (center) {LLM-assisted hardware\\testbench generation};
\node[diagramtask, below left=0.9cm and 0.2cm of center, minimum width=3.4cm] (stimulus) {Stimulus generation\\vectors, transactions,\\coverage-bin targets};
\node[diagramtask, below=0.9cm of center, minimum width=3.4cm] (bench) {Testbench construction\\clocks, resets, drivers,\\monitors, checkers};
\node[diagramtask, below right=0.9cm and 0.2cm of center, minimum width=3.4cm] (env) {Environment synthesis\\UVM agents, sequences,\\protocol libraries};
\node[diagramtool, below=1.05cm of stimulus, minimum width=3.0cm] (coverage) {Coverage feedback\\line, branch, FSM,\\functional bins};
\node[diagramtool, below=1.05cm of bench, minimum width=3.0cm] (oracle) {Oracle feedback\\scoreboards, mutants,\\bug injection};
\node[diagramtool, below=1.05cm of env, minimum width=3.0cm] (tools) {Tool feedback\\lint, compile, sim,\\formal checks};
\draw[diagramarrow] (center) -- (stimulus);
\draw[diagramarrow] (center) -- (bench);
\draw[diagramarrow] (center) -- (env);
\draw[diagramdash] (coverage) -- (stimulus);
\draw[diagramdash] (oracle) -- (bench);
\draw[diagramdash] (tools) -- (env);
\draw[diagramarrow] (stimulus) -- (bench);
\draw[diagramarrow] (bench) -- (env);
\node[draw, dashed, rounded corners, fit=(stimulus)(bench)(env)(coverage)(oracle)(tools), inner sep=7pt, label={[font=\scriptsize]below:Generation should be constrained by executable verification evidence, not only by natural-language plausibility.}] {};
\end{tikzpicture}
\caption{Taxonomy of LLM-assisted testbench generation tasks and the main verification signals used to constrain them.}
\label{fig:testbench-taxonomy}
\end{figure*}
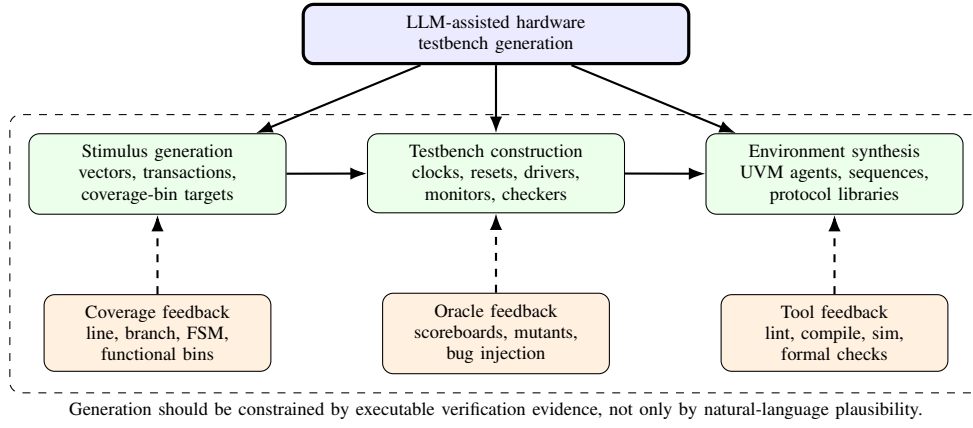

Figure~\ref{fig:testbench-taxonomy} summarizes this taxonomy and the feedback signals that cut across the three task levels.

\subsection{Pre-LLM Automation: Templates, Grammars, and Constrained Random Testing}

Automatic testbench generation predates modern LLMs.
Earlier tools demonstrate two lessons that remain important.
First, constraining the output space greatly improves syntactic stability.
Second, simulation throughput matters because verification requires repeated execution.
AVERT, for example, generates Verilog testbenches for grammatical-evolution hardware-design workflows by using DUT port information and standard test-vector files; it supports combinational and sequential circuits and improves simulation efficiency by evaluating multiple DUTs in a single generated testbench \cite{avert2022}.
Such approaches do not solve oracle discovery or semantic adequacy, but they show that structured generation can remove repetitive formatting work and reduce avoidable syntax errors.

Other pre-LLM work includes coverage-directed test generation, directed testing, and RTL fuzzing.
Machine-learning-assisted coverage-directed generation and directed hardware validation emphasizes the same optimization problem that LLM systems now revisit: selecting tests that close coverage efficiently while preserving bug-finding power \cite{ioannides2012cdg,jayasena2024directed}.
RFUZZ, DirectFuzz, and FuzzWiz show that fuzzing-style feedback can drive RTL exploration using coverage or hardware-specific guidance, making them important non-LLM baselines for stimulus-generation studies \cite{rfuzz2018,directfuzz2021,fuzzwiz2024}.

Traditional constrained-random verification similarly provides a useful baseline for LLM work.
Randomization and coverage feedback can explore large input spaces, especially in UVM-style environments, but engineers still have to write constraints, monitors, scoreboards, and coverage models.
LLM-based methods therefore should not be evaluated only against direct prompting; they should also be compared with random, constrained-random, template-based, fuzzing-based, and coverage-directed baselines where appropriate.
This is especially important because a generated testbench that improves line coverage may still miss a functional bug if the checker is weak or absent.

\subsection{Prompted LLM Generation with EDA Feedback}

The earliest LLM-based testbench studies generally prompt a commercial model with RTL and a natural-language specification, compile or simulate the result, and feed tool output back into subsequent prompts.
Bhandari et al. study this pattern for finite-state-machine testbench generation and bug detection.
Their workflow asks GPT-3.5 or GPT-4 to generate tests, uses commercial EDA feedback to identify uncovered transitions or errors, and iteratively refines prompts until coverage improves.
The work is important because it treats simulator and coverage output as first-class prompt context, not merely as an after-the-fact evaluation step \cite{bhandari2024fsm}.

The FSM study also exposes a recurring limitation: LLMs struggle with long traces, many states, and multi-bit outputs.
The authors mitigate this by chunking traces, resetting after groups of transitions, and decomposing multi-bit outputs into smaller subproblems.
This suggests a broader design principle for testbench generation: scalable systems should avoid asking the model to reason over an entire trace or full environment in one context window.
Instead, they should decompose the task into local coverage gaps, individual protocol obligations, or smaller signal cones.

\subsection{Coverage-Directed Stimulus Generation}

A second line of work uses the LLM primarily as a coverage-directed stimulus generator.
LLM4DV is an open-source framework that orchestrates LLMs for hardware test-stimuli generation across multiple DUTs.
Rather than asking for a complete verification environment, it iteratively samples missed coverage bins, restarts or refines dialogue context, and uses few-shot prompting to guide the generation of input stimuli \cite{llm4dv2023}.
This narrower framing is valuable because coverage bins and simulator feedback provide an executable objective: the model is rewarded for reaching untested behavior, not for producing aesthetically plausible HDL.

VerilogReader takes a related but more code-coverage-oriented view.
It integrates an LLM into a coverage-directed generation loop where the model acts as a ``Verilog reader'' that interprets RTL and coverage reports, then proposes stimuli for uncovered branches \cite{verilogreader2024}.
The system's Coverage Explainer and DUT Explainer modules illustrate a practical prompt-engineering pattern: raw Verilator output is not ideal LLM context, so an intermediate explanation layer translates simulator artifacts into natural-language goals.
The method performs best when the design remains within the model's comprehension scope, and its degradation on larger FSMs reinforces the scalability concerns observed in FSM-focused work.

Spec2Cov generalizes this agentic coverage-closure perspective by driving test-stimulus generation directly from design specifications and iteratively coordinating compilation, simulation, coverage-report parsing, and prompt refinement \cite{spec2cov2026}.
Its results show that agentic coverage closure can reach full coverage on simpler designs but remains significantly harder on complex designs, which is consistent with the broader literature: coverage feedback is powerful, but it does not eliminate the need for design understanding, good oracles, and scalable decomposition.

Several newer systems frame coverage closure as part of a broader verification-agent loop.
PRO-V-R1 combines an LLM with reinforcement learning and simulation feedback to improve automatic RTL verification, while UCAgent proposes an end-to-end block-level verification agent that traverses states and goals toward coverage closure \cite{prov2025,ucagent2026}.
These systems are not only testbench generators; they combine planning, test generation, execution, and refinement.
They are therefore best viewed as adjacent evidence that testbench generation is becoming an agentic workflow rather than a single prompt-completion task.

Figure~\ref{fig:feedback-loop} abstracts the recurring loop shared by these systems.

\begin{figure*}[t]
\centering
\begin{tikzpicture}[node distance=0.85cm and 0.7cm]
\node[diagramhead, minimum width=2.55cm] (spec) {Spec + RTL\\interfaces, resets,\\protocol intent};
\node[diagramtask, right=of spec, minimum width=2.55cm] (plan) {LLM planner\\scenarios, signals,\\coverage goals};
\node[diagramtask, right=of plan, minimum width=2.55cm] (gen) {Generator\\stimuli, drivers,\\checkers};
\node[diagramtool, right=of gen, minimum width=2.55cm] (exec) {EDA execution\\lint, compile,\\simulate, prove};
\node[diagramtool, below=1.05cm of exec, minimum width=2.7cm] (metrics) {Evidence\\errors, traces,\\coverage, mutants};
\node[diagramtask, left=of metrics, minimum width=2.7cm] (repair) {Repair/alignment\\prompt update,\\self-correction, DPO};
\node[diagramrisk, left=of repair, minimum width=2.7cm] (review) {Human review\\oracle adequacy,\\signoff judgment};
\draw[diagramarrow] (spec.east) -- (plan.west);
\draw[diagramarrow] (plan.east) -- (gen.west);
\draw[diagramarrow] (gen.east) -- (exec.west);
\draw[diagramarrow] (exec.south) -- (metrics.north);
\draw[diagramarrow] (metrics.west) -- (repair.east);
\draw[diagramarrow] (repair.north) -- (gen.south);
\draw[diagramdash] (review.north) -- (plan.south);
\draw[diagramdash] (review.east) -- (repair.west);
\node[font=\scriptsize, align=center, below=0.34cm of repair] {Solid arrows: automated tool-feedback loop. Dashed arrows: human constraints and review.};
\end{tikzpicture}
\caption{Verification-aware feedback loop used by simulator-in-the-loop, self-correcting, coverage-directed, and agentic testbench-generation systems.}
\label{fig:feedback-loop}
\end{figure*}

\subsection{Full Testbench Generation and Self-Correction}

AutoBench is a key transition from stimulus-only work to complete LLM-generated testbenches.
It proposes a staged workflow that separates scenario specification, Verilog driver generation, and Python checker generation.
This architecture deliberately uses Python for checking because LLMs are generally stronger at Python than HDL, while Verilog remains necessary for driving and integrating with RTL simulation \cite{autobench2024}.
AutoBench also introduces AutoEval, which evaluates generated testbenches at multiple levels: basic execution, functional behavior, and mutation-based adequacy.
The mutation-based step is particularly important because a testbench can compile and even pass the golden design while still failing to reject incorrect RTL mutants.

CorrectBench extends this direction by adding functional self-validation and self-correction.
Its central insight is that syntax-level debugging is insufficient: the generated checker or driver may be functionally wrong even after it compiles.
CorrectBench uses scenario-based validation with imperfect RTL variants to infer whether a generated testbench can distinguish correct and incorrect behavior, then feeds detected bug information to an LLM-based corrector \cite{correctbench2025}.
This shifts the target from ``does the testbench run?''
to ``does the testbench detect meaningful behavioral differences?''

ConfiBench further extends the AutoBench/CorrectBench family with confidence-based scenario masking and a multi-testbench ensemble.
Instead of trusting a single generated testbench in all scenarios, it masks high-risk scenarios and combines multiple generated testbenches to compensate for coverage loss.
Reported results show a higher overall pass rate than AutoBench and CorrectBench-style baselines, especially on sequential circuits \cite{confibench2025}.
The broader lesson is that ensemble and confidence mechanisms can be useful when individual LLM outputs remain unstable.

\subsection{UVM and Subsystem-Level Generation}

Moving from module-level Verilog testbenches to UVM environments introduces a substantial jump in complexity.
UVM code is class-heavy, convention-driven, and protocol-sensitive; a generated environment must coordinate transaction definitions, sequences, drivers, monitors, scoreboards, configuration databases, coverage groups, and simulator-specific build rules.
VerifLLMBench addresses this evaluation gap by benchmarking LLM-generated UVM testbenches using verified DUTs, generated UVM components, iterative syntax repair, linting, compilation, simulation, and coverage analysis \cite{verifllmbench2025}.
Its importance is methodological: UVM generation must be evaluated as an environment-construction problem, not merely as a single-file code-generation task.

UVM\textsuperscript{2} demonstrates a more automated UVM workflow that generates UVM testbenches and refines stimuli with coverage feedback on RTL designs up to roughly 1.6K lines of code \cite{uvm2_2025}.
HAVEN pushes the structured-generation principle further by preventing the LLM from writing raw UVM code directly.
Instead, LLM agents extract an architectural plan, protocol-specific Jinja2 templates instantiate UVM components, and a protocol-aware sequence DSL generates bus-correct stimulus.
On open-source IPs spanning Direct, Wishbone, and AXI4-Lite interfaces, HAVEN reports complete compilation success and strong average code and functional coverage \cite{haven2026}.
This is one of the clearest examples that reducing LLM freedom can improve verification trust.

UVMarvel extends the UVM story to subsystem-level verification.
It introduces an intermediate representation, a bus protocol library, signal tracking, and a Verilog patching library to translate heterogeneous specifications into protocol-correct UVM environments and refine stimuli \cite{uvmarvel2026}.
Compared with module-level benchmarks, this is closer to industrial verification because subsystems involve protocol interactions, generated interconnect, and multiple coordinated blocks.
However, it also shows that successful LLM-aided UVM generation currently depends on substantial handcrafted infrastructure: protocol libraries, IRs, templates, and tool-specific repair loops.

Other UVM-oriented and end-to-end verification studies support the same trend.
Exploratory APB UVM case studies show that generative models can help scaffold protocol-specific UVM components, but also that human review is still needed for completeness and correctness \cite{dranga2024apbuvm}.
Multi-agent verification frameworks and security-verification agents extend the idea beyond plain functional testbenches by combining LLM-generated collateral with simulator execution, assertion/security checks, or tool-guided debug loops \cite{mavf2025,svllm2025,promptverifyrepeat2025}.
These works are useful related literature even when their primary target is broader verification automation rather than testbench generation alone.

\subsection{Training-Time Alignment and Execution-Aware Learning}

Prompting and agentic refinement improve inference-time behavior, but they do not directly teach the model which testbench patterns produce high verification value.
TB or Not TB addresses this by fine-tuning models with Coverage-Driven Direct Preference Optimization (CD-DPO).
It constructs PairaNet, a preference dataset of higher- and lower-quality testbenches labeled by simulation-derived coverage metrics, and trains models to prefer stimuli that improve coverage \cite{tbornottb2025}.
This is important because it turns coverage from a prompt-time hint into a training signal.

LLM4Cov similarly argues that online reinforcement learning with simulators is often too slow and expensive, so it builds an offline execution-aware agent-learning framework.
The framework uses execution-validated data curation, policy-aware synthesis, and worst-state-prioritized sampling to train compact models for high-coverage testbench generation \cite{llm4cov2026}.
These training-time approaches indicate a likely future direction: instead of relying only on increasingly large general models, verification-specific generators may be trained on simulator-labeled examples, coverage traces, failed repair attempts, and tool-grounded preferences.

Structured Testbench Generation (STG) takes an even stronger stance against unconstrained LLM generation.
Rather than iteratively prompting an LLM to synthesize testbenches, STG exploits hardware structure to generate deterministic testbenches, reduce false passes, and curate verification-oriented data for model training \cite{stg2026}.
STG's results support a recurring conclusion across this section: testbench generation benefits from deterministic structure wherever possible, with LLMs used for planning, classification, repair, or data augmentation rather than unrestricted final-code authorship.

\subsection{Agentic Verification Workflows}

The newest papers increasingly describe testbench generation as one component of an agentic verification loop.
In these systems, the LLM may decompose a verification goal, generate or repair testbench code, execute tools, interpret compiler or simulator output, update coverage goals, and repeat.
This matches the ``prompt, verify, repeat'' pattern articulated in recent verification-cycle work \cite{promptverifyrepeat2025}.
The important distinction from earlier self-debugging is that the agent owns more of the verification workflow: it can select which scenario to target, which tool output to inspect, and when to revise code or stimuli.

This agentic framing has two consequences for a survey.
First, related work should include papers whose output is not only a testbench file but a verification process that includes testbench generation.
PRO-V-R1, UCAgent, MAVF, and SV-LLM are examples because they use generated programs, tests, or verification collateral inside larger automatic RTL verification pipelines \cite{prov2025,ucagent2026,mavf2025,svllm2025}.
Second, evaluation needs to include cost and convergence, not only final coverage: the number of compile/simulation attempts, failed repairs, prompts, and tool calls becomes part of the method's practicality.

\subsection{Benchmarks and Evaluation Metrics}

A serious testbench-generation benchmark must evaluate several layers of correctness.
At minimum, the generated artifact should compile, elaborate, and simulate.
However, these checks are only gateway metrics.
A stronger benchmark asks whether the testbench drives meaningful scenarios, reaches structural and functional coverage targets, detects injected bugs or RTL mutants, and avoids false passes.
AutoBench and CorrectBench make this explicit through multi-stage evaluation and mutant-based checks \cite{autobench2024,correctbench2025}.
CVDP broadens the benchmark landscape by including hardware design and verification tasks in both non-agentic and agentic forms, revealing that tool-using verification tasks remain difficult for current models \cite{cvdp2025}.
Benchmarking should also include fuzzing and directed-test baselines because classical CDG, RFUZZ, DirectFuzz, and FuzzWiz attack the same coverage-closure objective with non-LLM search strategies \cite{ioannides2012cdg,rfuzz2018,directfuzz2021,fuzzwiz2024}.

The most useful metrics for this subfield include:
\begin{itemize}[leftmargin=1.5em]
    \item \textbf{Build success}: parsing, elaboration, simulator compatibility, and lint cleanliness;
    \item \textbf{Stimulus effectiveness}: line, branch, toggle, FSM-state, FSM-transition, and user-defined functional coverage;
    \item \textbf{Oracle strength}: ability to reject faulty designs, RTL mutants, or injected bugs;
    \item \textbf{Repair convergence}: number of tool-feedback iterations, token cost, and simulator runs required;
    \item \textbf{Scalability}: performance on sequential circuits, larger FSMs, bus protocols, and subsystem-level designs;
    \item \textbf{Portability}: whether generated collateral works across open-source and commercial simulators.

\end{itemize}

Figure~\ref{fig:evaluation-ladder} shows why later-stage metrics provide stronger evidence than syntax or compilation alone.
Tables~\ref{tab:testbench-related-work-foundations}--\ref{tab:testbench-related-work-agentic} summarize representative work across these dimensions.

\begin{figure*}[t]
\centering
\begin{tikzpicture}[node distance=0.28cm]
\node[diagramnode, fill=gray!10, minimum width=2.25cm] (build) {1. Build\\parse, elaborate, lint};
\node[diagramnode, fill=blue!7, right=of build, minimum width=2.25cm] (sim) {2. Simulation\\expected scenarios};
\node[diagramnode, fill=green!8, right=of sim, minimum width=2.25cm] (cov) {3. Coverage\\branches, FSMs, bins};
\node[diagramnode, fill=orange!12, right=of cov, minimum width=2.25cm] (oracle) {4. Oracle\\mutants, bugs, checks};
\node[diagramnode, fill=red!8, right=of oracle, minimum width=2.25cm] (port) {5. Readiness\\portability, review};
\draw[diagramarrow] (build.east) -- (sim.west);
\draw[diagramarrow] (sim.east) -- (cov.west);
\draw[diagramarrow] (cov.east) -- (oracle.west);
\draw[diagramarrow] (oracle.east) -- (port.west);
\node[font=\scriptsize, below=0.35cm of cov, align=center] {Evidence strength increases from left to right; later stages are harder to satisfy but more meaningful for signoff.};
\end{tikzpicture}
\caption{Evaluation ladder for generated testbenches. Compilation is necessary, but coverage, oracle strength, mutation resistance, and portability are stronger indicators of verification value.}
\label{fig:evaluation-ladder}
\end{figure*}

\begin{table*}[t]
\centering
\caption{Representative related work on automatic and LLM-assisted hardware testbench generation: classical baselines and stimulus-generation methods.}
\label{tab:testbench-related-work-foundations}
\tiny
\setlength{\tabcolsep}{2pt}
\renewcommand{\arraystretch}{1.1}
\begin{tabularx}{\textwidth}{P{1.8cm}YYYYY}
\toprule
\textbf{Work} & \textbf{Primary target} & \textbf{Generation style} & \textbf{Feedback / evaluation} & \textbf{Main contribution} & \textbf{Key limitation} \\
\midrule
AVERT \cite{avert2022} & Verilog testbench construction for evolutionary hardware & Grammar/template-based generation from ports and test vectors & Simulation throughput on benchmark circuits & Shows that structured non-LLM generation can reduce manual testbench setup and improve simulation efficiency & Requires externally supplied vectors and does not infer semantic oracles \\
Bhandari et al. \cite{bhandari2024fsm} & FSM testbench generation and bug detection & Prompted GPT-3.5/GPT-4 with iterative EDA feedback & Transition coverage and bug-detection traces & Demonstrates compiler/simulator feedback as a practical prompt-refinement signal & Scalability limited by trace length, state count, and context management \\
RFUZZ \cite{rfuzz2018} & RTL fuzzing and coverage exploration & Coverage-guided fuzzing for RTL simulators & Coverage feedback and bug-finding experiments & Establishes fuzzing as a strong non-LLM coverage-directed baseline & Does not generate human-readable verification environments or semantic checkers \\
DirectFuzz \cite{directfuzz2021} & Directed RTL test generation & Directed fuzzing with hardware-aware guidance & Targeted coverage and bug exposure & Improves reachability of hard-to-trigger RTL behaviors & Still needs external checking and target selection \\
FuzzWiz \cite{fuzzwiz2024} & RTL fuzzing with hardware-specific feedback & Automated hardware fuzzing workflow & Coverage and crash/bug finding & Provides a recent fuzzing baseline for automated RTL stimulus generation & Optimizes exploration more than reusable testbench construction \\
LLM4DV \cite{llm4dv2023} & Hardware test-stimuli generation & Coverage-directed prompting and missed-bin refinement & Functional coverage across multiple DUTs & Frames LLMs as stimulus generators inside a DV loop & Less focused on complete reusable testbench environments \\
VerilogReader \cite{verilogreader2024} & Code-coverage-directed stimulus generation & LLM reads RTL plus coverage explanations & Verilator-style coverage and random-testing comparison & Introduces coverage and DUT explainers to translate tool output into LLM-readable goals & Degrades when designs exceed model comprehension scope \\
Spec2Cov \cite{spec2cov2026} & Specification-driven coverage closure & Agentic prompt refinement with simulator interaction & Compilation, simulation, and code coverage over varied designs & Automates iterative stimulus generation from specifications & Full coverage remains difficult for complex designs \\
\bottomrule
\end{tabularx}
\end{table*}

\begin{table*}[t]
\centering
\caption{Representative related work on automatic and LLM-assisted hardware testbench generation: full testbench, UVM, and subsystem-level methods.}
\label{tab:testbench-related-work-uvm}
\tiny
\setlength{\tabcolsep}{2pt}
\renewcommand{\arraystretch}{1.1}
\begin{tabularx}{\textwidth}{P{1.8cm}YYYYY}
\toprule
\textbf{Work} & \textbf{Primary target} & \textbf{Generation style} & \textbf{Feedback / evaluation} & \textbf{Main contribution} & \textbf{Key limitation} \\
\midrule
AutoBench \cite{autobench2024} & Full Verilog/Python hybrid testbench generation & Scenario, driver, and checker decomposition & Eval0--Eval2 metrics with mutant RTLs & Establishes a systematic LLM testbench-generation and evaluation framework & Syntax-focused repair still leaves functional checker errors \\
CorrectBench \cite{correctbench2025} & Functionally self-correcting testbench generation & Generator, validator, and corrector agents & Scenario validation and Eval2 pass ratio & Adds functional self-validation and self-correction without golden testbenches & Still relies on imperfect generated variants and benchmark-scale tasks \\
ConfiBench \cite{confibench2025} & Robust testbench generation under uncertainty & Scenario masking and multi-testbench ensemble & Pass rate on AutoBench-style tasks & Uses confidence and ensembles to reduce risky generated behavior & Added complexity and cost; still task-benchmark centered \\
VerifLLMBench \cite{verifllmbench2025} & UVM testbench benchmark & LLM-generated UVM components with iterative repair & Build success, lint, simulation, and coverage & Provides an open methodology for evaluating LLM-generated UVM environments & Demonstration designs are relatively simple compared with industrial subsystems \\
UVM\textsuperscript{2} \cite{uvm2_2025} & UVM testbench and stimulus generation & Automated UVM framework with coverage-guided refinement & Code and functional coverage on larger RTL designs & Moves LLM verification from simple benches toward reusable UVM workflows & Requires substantial tool orchestration and domain-specific setup \\
HAVEN \cite{haven2026} & UVM testbench and sequence synthesis & LLM planning plus protocol templates and sequence DSL & Compilation, code coverage, and functional coverage & Avoids raw UVM generation by using templates and protocol-aware DSLs & Protocol-specific engineering must be built and maintained \\
UVMarvel \cite{uvmarvel2026} & Subsystem-level UVM generation & IR, protocol library, signal tracking, and patching & Subsystem code coverage and automated runtime & Extends LLM-aided UVM generation to subsystem-level protocols & Public evidence remains recent and protocol-library dependent \\
\bottomrule
\end{tabularx}
\end{table*}

\begin{table*}[t]
\centering
\caption{Representative related work on automatic and LLM-assisted hardware testbench generation: training-time and agentic methods.}
\label{tab:testbench-related-work-agentic}
\tiny
\setlength{\tabcolsep}{2pt}
\renewcommand{\arraystretch}{1.1}
\begin{tabularx}{\textwidth}{P{1.8cm}YYYYY}
\toprule
\textbf{Work} & \textbf{Primary target} & \textbf{Generation style} & \textbf{Feedback / evaluation} & \textbf{Main contribution} & \textbf{Key limitation} \\
\midrule
TB or Not TB \cite{tbornottb2025} & Coverage-optimized stimulus generation & CD-DPO fine-tuning from coverage-labeled preferences & CVDP CID12 coverage metrics & Uses simulator-derived coverage as a training-time preference signal & Depends on curated paired data and simulator-labeled examples \\
LLM4Cov \cite{llm4cov2026} & High-coverage agentic testbench generation & Offline execution-aware agent learning & Coverage pass rate under agentic evaluation & Learns from execution-validated trajectories without online RL cost & Benchmark and training protocol are recent; industrial generality remains open \\
PRO-V-R1 \cite{prov2025} & Automatic RTL verification agent & LLM plus reinforcement learning and simulation feedback & Program/test generation and verification outcomes & Treats verification as an executable feedback loop rather than static generation & Broader RTL-verification scope makes direct TB-only comparison difficult \\
UCAgent \cite{ucagent2026} & End-to-end block-level functional verification & Agentic state/goal exploration with tool feedback & Coverage closure and verification progress & Shows how test generation can be embedded in a full verification agent & Very recent preprint; benchmark generality remains evolving \\
MAVF \cite{mavf2025} & Module-level verification automation & Multi-agent generative AI workflow & Verification artifact generation and tool execution & Demonstrates division of labor among agents for module-level verification & Industry-style framework with limited public reproducibility details \\
SV-LLM \cite{svllm2025} & SoC security verification & Agentic LLM workflow for security checks and generated collateral & Security-verification outcomes & Extends LLM-generated verification artifacts to security-oriented SoC flows & Security focus differs from general functional testbench generation \\
STG \cite{stg2026} & Deterministic testbench generation and data curation & Structure-driven deterministic generation & Runtime, compile success, coverage, and false-pass reduction & Shows deterministic structure can outperform iterative LLM testbench synthesis in speed and reliability & Less flexible when verification intent is not recoverable from structure alone \\
\bottomrule
\end{tabularx}
\end{table*}

%
\subsection{Challenges and Future Directions}
\label{sec:tb-challenges}

LLM-based testbench generation has evolved from generating directed tests for small RTL modules toward self-checking testbenches, coverage-guided refinement, UVM environments, and subsystem-level verification. 
In parallel, workflows have progressed from one-shot generation toward iterative and agentic approaches that incorporate simulation and coverage feedback. 
Despite this progress, unconstrained SystemVerilog and UVM generation remains unreliable for complex sequential and protocol-intensive designs. 
The strongest approaches therefore combine LLM reasoning with deterministic structures such as templates, intermediate representations (IRs), domain-specific languages (DSLs), protocol libraries, and verification-tool feedback.

A central challenge is \textbf{oracle quality}. Effective stimuli provide limited value without trustworthy mechanisms for determining correctness. High coverage or successful execution against golden RTL does not necessarily imply strong bug-detection capability. Mutation testing, imperfect RTL variants, and controlled bug injection therefore provide important complementary evaluation~\cite{autobench2024,correctbench2025,stg2026}. Future work should place greater emphasis on generating and validating scoreboards, reference models, assertions, and other checking mechanisms.

\textbf{Sequential and temporal reasoning} remains difficult for resets, handshakes, long-latency operations, pipelines, FSMs, and subsystem protocols. Existing approaches mitigate this challenge by decomposing behavior into scenarios, transactions, DSL operations, or targeted coverage gaps~\cite{bhandari2024fsm,verilogreader2024,haven2026}. Extending such structured reasoning to longer traces and interacting protocols is important for scaling beyond small designs.

\textbf{Industrial-scale UVM generation} introduces additional challenges because useful environments must correctly connect components, obey protocol timing, implement meaningful scoreboards and coverage models, and integrate with simulator workflows. HAVEN, UVM\textsuperscript{2}, UVMarvel, and VerifLLMBench demonstrate progress, but also indicate that UVM generation benefits substantially from templates, IRs, and reusable protocol libraries~\cite{verifllmbench2025,uvm2_2025,haven2026,uvmarvel2026}. This suggests a hybrid approach in which LLMs perform high-level planning and reasoning while deterministic mechanisms generate structurally constrained verification components.

\textbf{Scalability and evaluation} also remain open problems. HDLBits- and VerilogEval-derived tasks provide reproducibility but are generally too small and public to establish industrial readiness and may introduce contamination concerns. CVDP and UVM-oriented benchmarks improve realism~\cite{cvdp2025,verifllmbench2025}, but broader evaluation is needed across hidden designs, large subsystems, mixed-language repositories, multiple simulators, and standardized bug suites. Richer datasets aligning specifications, RTL, testbenches, execution traces, coverage, and bugs would further support both evaluation and verification-aware training.

Finally, \textbf{execution-driven refinement can be costly}. Repeated LLM and simulator interactions increase inference cost, simulator usage, and verification time. Promising directions include learning from execution traces, deterministic structural generation, targeted repair, and agentic controllers that obtain more useful feedback from fewer tool runs~\cite{tbornottb2025,llm4cov2026,prov2025,ucagent2026,stg2026}. Evaluation should therefore report simulator runs, refinement iterations, inference cost, and human intervention in addition to functional results.

Overall, existing work suggests that progress in LLM-based testbench generation depends less on unconstrained code-generation capability or model scale than on how effectively the LLM is integrated with the verification process. 
The field is consequently shifting from prompt-based testbench generation toward \emph{verification-aware automation}, in which simulator outcomes serve as validation signals, refinement feedback, or training preferences, while deterministic structures constrain generation. 
Future progress will likely depend on stronger oracle generation, improved temporal and protocol reasoning, richer specification--RTL--testbench datasets, simulator-portable and contamination-resistant evaluation, and scalable workflows that combine LLM planning with deterministic verification infrastructure.
\section{Debugging and Design Repair}
\label{sec:debug}

%
Sections~III and IV examined how LLMs can assist pre-silicon verification by generating assertions, stimuli, testbenches, and other verification artifacts to expose incorrect RTL behavior. 
Once a failure is detected, however, compiler diagnostics, simulation mismatches, assertion violations, or formal counterexamples must be translated into an actionable diagnosis and a validated correction. 
This motivates \emph{debugging and design repair}, which closes the verification loop by identifying the fault, determining its cause, correcting the RTL, and re-verifying the design.

\begin{figure*}
    \centering
    \includegraphics[width=\textwidth]{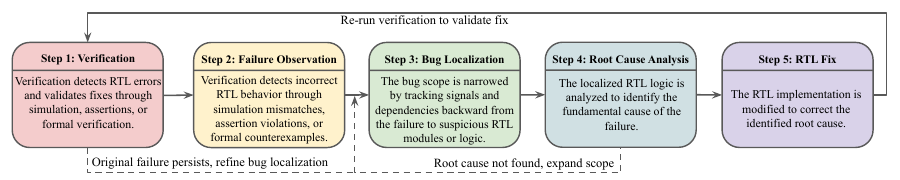}
    \caption{Conventional RTL debugging process illustrating iterative bug
    localization, root cause analysis, RTL fixing, and re-verification.}
    \label{fig:debug_process}
\end{figure*}

As illustrated in Fig.~\ref{fig:debug_process}, RTL debugging is an iterative failure-to-repair process. 
\textbf{Bug localization} narrows an observed failure to suspicious RTL modules, blocks, expressions, signals, or assignments using the RTL and failure evidence such as waveforms, assertion reports, or formal counterexamples~\cite{bg_pal,dbg_lik,dbg_hdldebugger}.
\textbf{Root cause analysis (RCA)} determines how the localized logic produces the observed deviation, while \textbf{RTL repair} generates a corrective modification for re-verification~\cite{bg_witharana}. 
If the failure persists or new violations emerge, the relevant stages are repeated. 
These stages should not be conflated: accurate localization does not establish the root cause, a plausible diagnosis does not guarantee a correct repair, and a patch that passes the available checks may not preserve behavior outside those checks. 
Source-only localization further identifies suspicious RTL without an
observed failure, whereas assertion repair corrects the verification property rather than the RTL implementation~\cite{dbg_flag,dbg_lik,dbg_assertfix}.
Manual execution of this loop remains a major verification bottleneck~\cite{bg_pal}, increasingly aggravated by design scale and
integration complexity~\cite{bg_witharana}.

RTL concurrency and hierarchy make this process particularly challenging. 
A defect may propagate across multiple modules and clock cycles before becoming observable, making the reported failure structurally and temporally distant from its source~\cite{bg_pal,dbg_hdldebugger}. 
Repair must further relate the suspected logic to the specification and timing behavior, remove the failure, and preserve unaffected functionality. 
Errors in localization or diagnosis can therefore propagate to invalid or incomplete repairs and additional debugging iterations~\cite{bg_witharana}.

LLMs can support this workflow by jointly processing RTL, specifications,
diagnostics, simulation traces, and formal counterexamples to assist
localization, causal diagnosis, and repair~\cite{dbg_hdldebugger,dbg_lik,
dbg_fvdebug}. 
Hardware-specific methods strengthen these capabilities through RAG for external design or repair knowledge~\cite{dbg_rtlfixer}, RTL-specific
fine-tuning~\cite{dbg_hdldebugger,dbg_assertsolver}, explicit dependency
representations~\cite{dbg_fvdebug}, and multi-agent coordination of
localization and repair~\cite{dbg_meic,dbg_r3a}. 
These techniques provide complementary mechanisms for implementing the failure-to-repair workflow.

In this section, we survey 25 papers on LLM-assisted hardware debugging, encompassing bug detection and localization, RCA, RTL and assertion repair, and integrated debugging workflows. 
Section~\ref{subsec:debug-methods} classifies the works by their primary debugging methodology, while Section~\ref{subsec:debug-pipeline} compares their coverage of the debugging pipeline. Section~\ref{subsec:debug-generation} examines debugging embedded within RTL generation workflows. 
Section~\ref{subsec:self-evolving-debug} concludes with open challenges and future directions. 
General-purpose software-repair methods without hardware evaluation are excluded.
%
\subsection{LLM-Based Debugging Methodologies Taxonomy}
\label{subsec:debug-methods}

LLM-assisted debugging methods span multiple stages of the debugging pipeline but differ in how they acquire and use information to guide localization, diagnosis, and repair. 
To characterize these methodological differences, we group the reviewed works into four categories according to their principal mechanism: (1) prompting and tool-feedback-based debugging; (2) retrieval-augmented debugging; (3) model adaptation; and (4) structured and agentic reasoning. 
These categories capture complementary sources of knowledge and control, enabling comparison of their capabilities, requirements, and limitations beyond the pipeline stages they address.
The categories are not mutually exclusive or indicative of capability.
Many frameworks combine multiple mechanisms; each work is categorized by its principal contribution, with additional mechanisms identified in Tables~\ref{tab:debug-part1} and~\ref{tab:debug-part2}.

\subsubsection{Category 1: Prompting and Tool-Feedback-Based Debugging}

This category uses prompting to guide LLMs in identifying defects or generating candidate repairs. Some methods regenerate RTL and analyze differences from the original code to locate potential defects \cite{dbg_flag}. 
Others incorporate compiler or simulation feedback to iteratively refine repairs \cite{dbg_meic}. These approaches primarily rely on knowledge encoded in the LLM and information provided in the prompt, without external retrieval or model parameter updates.

Prompting methods can inject task-specific knowledge without modifying model parameters. 
Ahmad et al.~\cite{dbg_securityfix} investigate repair of
human-introduced hardware security bugs using CWE-conditioned prompts and
model/prompt sampling. Across 15 bugs, every case has at least one successful
repair candidate, while GPT-4 produces 3,862 correct repairs in 7,500 attempts.
For selected weaknesses, CWEAT is used for bug detection and post-repair
checking. \textit{Self-HWDebug}~\cite{dbg_selfhwdebug} instead derives reusable
repair instructions from vulnerable/secure RTL pairs. Its two-shot
configuration repairs all 25 test snippets spanning five CWE categories.
Both approaches encode vulnerability knowledge through prompts rather than
external retrieval or model adaptation. Beyond repair, prompting can support bug detection and localization. \textit{FLAG}~\cite{dbg_flag} prompts an LLM to regenerate RTL and uses differences from the original code, token confidence, and distance from comments as features for defect localization. Because it does
not require a failing testbench, it can screen incomplete or non-compiling
designs. Its line-level method detects 38 of 120 OpenTitan bugs, demonstrating how prompted RTL regeneration can provide diagnostic features for source-level screening. 

Tool-feedback-based methods close the loop by using compiler or simulation results to evaluate generated repairs and guide subsequent revisions. \textit{MEIC}~\cite{dbg_meic} employs debugging and scoring agents, together with candidate history, rollback, and self-planning, to manage iterative repair. On 178 RTLLM-derived error instances, it reports syntax and functional fix rates of 93\% and 78\%, respectively. \textit{UVLLM}~\cite{dbg_uvllm} further combines
UVM checking with mismatch-driven data-flow slicing and test-score-based
rollback. Time-aware slicing narrows the failure context presented to the LLM,
while executable UVM checks guide candidate selection. Across 331 error
instances from 27 RTLLM-derived modules, it reports syntax and functional fix
rates of 86.99\% and 71.92\%, respectively.

These methods demonstrate that prompting and verification feedback can support
both source-level detection and iterative RTL repair without model adaptation.
Their effectiveness, however, depends strongly on the information exposed by
the prompt and verification tools. Ambiguous diagnostics or weak functional
feedback may not reveal the underlying cause, causing repeated generation to
converge slowly or fail when the required hardware knowledge is unavailable.
This limitation motivates retrieval-based methods that explicitly augment the
LLM with external debugging knowledge.

\subsubsection{Category 2: Retrieval-Augmented Debugging}

Retrieval-augmented methods supplement the LLM with external information, such as specifications, expert repair examples, documentation, or design context, at inference time.
This provides task-specific knowledge without modifying model parameters and can improve both diagnosis and repair when the required information is not reliably encoded in the LLM.

Qayyum et al.~\cite{dbg_bugsfixes} use RAG to retrieve specification context for HDL bug identification and repair.
Their method combines specification retrieval with expression decomposition and repairs all 25 nonconstant mutations across three OpenTitan IPs, but none of the five constant mutations.
The failed constants are absent from the specifications, illustrating that retrieval effectiveness depends on the completeness of the underlying knowledge source.

Rather than retrieving specifications, \textit{RTLFixer}~\cite{dbg_rtlfixer} retrieves validated expert solutions associated with compiler-error patterns and combines them with ReAct-style prompting and compiler feedback.
The framework reports a 98.5\% compilation-error repair rate, with ablations showing complementary contributions from retrieval, iterative feedback, and diagnostic quality.
\textit{GROVE}~\cite{dbg_grove} further makes the retrieval corpus adaptive by organizing previously resolved cases into an evolving knowledge tree with applicability conditions.
Candidate knowledge is validated on its originating cases, while budget-aware retrieval selects relevant entries for new failures.
On a module-disjoint SVA-Eval split, \textit{GROVE} improves Llama-3.3 pass@1 from 72.7\% to 84.7\% and achieves 86.9\% pass@5.
These results demonstrate the potential of reusable repair knowledge, although validation on an originating case does not guarantee transfer to other designs.

Retrieval can also select relevant design structure rather than textual repair knowledge.
\textit{HDLxGraph}~\cite{dbg_hdlxgraph} combines AST hierarchy with data-flow traversal to retrieve repository-level RTL context.
Its debugging evaluation covers five \texttt{mor1kx} commits and uses ROUGE rather than functionally validated repair, so the results demonstrate context retrieval rather than repair correctness.
\textit{ViTAD}~\cite{dbg_vitad} instead augments signal dependencies with timing information and retrieves repair strategies conditioned on inferred causes of timing violations.
Across 54 injected timing cases, \textit{ViTAD} improves the fix rate from 54.38\% to 71.05\%, a gain of 16.67 percentage points.

Overall, retrieval-augmented debugging depends on the relevance and completeness of external knowledge and the ability of the LLM to apply it correctly. 
\subsubsection{Category 3: Debugging through Model Adaptation}

Model adaptation encodes hardware-specific debugging knowledge in model parameters through continued pretraining, supervised fine-tuning, preference optimization, or specialized training objectives. These methods aim to improve localization and repair, but their effectiveness depends on representative training data. Public datasets pairing faulty RTL with fault locations, diagnostics, and validated repairs remain limited.

\textit{HDLdebugger}~\cite{dbg_hdldebugger} addresses this data limitation by constructing a synthetic debugging corpus through reverse engineering.
Controlled mutations of correct RTL are paired with compiler diagnostics and reference implementations to provide repair supervision, while code and document retrieval supplies additional context at inference time.
The resulting framework reports an 81.93\% execution pass rate on its test split drawn from 92,143 industrial HDL instances.
\textit{LLM4SecHW}~\cite{dbg_llm4sechw} instead mines open-source hardware version histories for defect/fix pairs and descriptions and applies lightweight adapter-based fine-tuning.
Its evaluation uses per-case ROUGE scores and qualitative examples rather than an aggregate functionally validated repair rate.
These approaches illustrate two complementary sources of adaptation data: controlled synthetic defects and naturally occurring development histories.

Model adaptation can also specialize LLMs for fault localization.
\textit{LiK}~\cite{dbg_lik} combines continual pretraining, supervised localization training, and SimPO preference optimization, achieving 93.38\% localization pass@1 compared with 77.9\% for GPT-4-Turbo on the same benchmark.
Supplying its predicted locations to GPT-3.5 further improves downstream repair from 40.39\% to 58.92\%, demonstrating the value of localization as an intermediate debugging step.
\textit{VeriDebug}~\cite{dbg_veridebug} instead integrates localization and repair within a shared model using contrastive representations of fault locations and bug types.
It reports 67.2\% top-1 localization accuracy and 64.7\% top-1 repair accuracy, where repair requires an exact match to the reference corrected line.
Thus, \textit{LiK} and \textit{VeriDebug} represent different interfaces between learned localization and repair: explicit localization for a downstream repair model versus joint localization and correction.

Adaptation can further target repair generation or constrain its scope.
\textit{AssertSolver}~\cite{dbg_assertsolver} combines pretraining, supervised fine-tuning, and direct preference optimization over correct and incorrect responses for assertion-triggering RTL faults.
It reports repair pass@1 and pass@5 of $88.54\%$ and $90.00\%$, respectively, and generates explanations alongside candidate corrections.
\textit{ARSP}~\cite{dbg_arsp} instead trains semantic partitioning and fragment-repair models to restrict modification to a selected RTL region while retaining the specification and global code as context.
On injected faults in 33 designs, it achieves $77.92\%$ pass@1 and $83.88\%$ pass@5.
These methods address complementary aspects of repair: response-level supervision improves candidate generation, whereas learned partitioning constrains where modifications are applied.

Overall, model adaptation enables hardware-specific localization and repair knowledge to be internalized rather than repeatedly retrieved or supplied through prompts.
The reviewed methods show that adaptation can improve localization, guide downstream repair, and specialize models for particular defect classes, while synthetic data, repository histories, and preference signals provide different forms of supervision.
However, performance depends strongly on the quality and representativeness of the training data, and metrics such as exact patch match or textual similarity do not necessarily establish functional correctness.
Moreover, parameter adaptation alone does not resolve failures whose causes span large RTL hierarchies, multiple clock cycles, or long verification traces.
These limitations motivate the structured representations, explicit search, and agent coordination examined in the next category.

\subsubsection{Category 4: Structured and Agentic Debugging}

Prompting, retrieval, and model adaptation improve the knowledge available to an LLM, but complex RTL failures may still require reasoning across multiple signals, modules, clock cycles, and candidate repairs.
Structured and agentic methods address this challenge by explicitly organizing failure evidence, intermediate hypotheses, repair search, or tool interactions.
This category therefore includes graph-based reasoning, structured search, and agent coordination, with classification determined by the principal methodological contribution rather than the use of multiple agents alone.

\textit{FVDebug}~\cite{dbg_fvdebug} structures formal counterexamples as a directed acyclic graph that captures signal dependencies across timesteps.
A Graph Scanner uses batched LLM inference with paired for-and-against prompting to identify suspicious nodes, after which a reasoning agent develops causal hypotheses and generates patches.
On~$38$ SVA-Eval-Human cases, \textit{FVDebug} reports a Quality@Best score of $0.956$ under an LLM judge and repair pass@1/pass@5 of $71.1\%$/$86.8\%$.
Its additional CVA6 and proprietary industrial examples primarily demonstrate RCA on larger designs and should not be interpreted as achieving the same repair rates.
\textit{VeriPilot}~\cite{dbg_veripilot} instead aligns RTL variables with internal variables of a supplied golden model and traces control- and data-flow dependencies to guide iterative repair.
On adapted CVDP tasks, this approach improves GPT-4o repair success from $54.3\%$ to $85.71\%$, demonstrating the value of explicit reference alignment while requiring an executable golden model.

\textit{VeriTrace}~\cite{dbg_veritrace:iclad26} introduces \emph{Agentic Temporal Exploration}, in which an Inspector agent iteratively selects internal signals and time windows for hypothesis-driven waveform analysis before a separate RTL Debugger performs targeted repair.
The Inspector reasons over the specification, RTL, and waveform evidence without direct access to the testbench, while candidate repairs are re-simulated against the supplied golden testbench and retained when they reduce mismatches.
On VerilogEval-V2, \textit{VeriTrace} achieves $100\%$ with Claude Sonnet 4.5, with notable gains on sequential FSM tasks.
However, repair acceptance is determined by the supplied testbench, so passing establishes consistency with the exercised behaviors rather than independent conformance to the complete specification; behaviors, corner cases, or temporal conditions omitted by the testbench remain unverified.

Structured reasoning can also organize the search over candidate repairs.
\textit{R3A}~\cite{dbg_r3a} separates multi-agent fault localization from patch exploration, using ranked candidate locations to initialize a stochastic tree-of-thoughts search.
A heuristic balances exploration and exploitation under a fixed time budget, enabling \textit{R3A} to solve $29$ of $32$ RTL-Repair cases, with mean pass@1 and pass@5 of $75.0\%$ and $86.7\%$, respectively.
\textit{Clover}~\cite{dbg_clover} extends this approach by dispatching repair tasks between LLM-based editing and SMT-based repair within a dialogue-state tree.
It solves $31$ of $32$ RTL-Repair cases and achieves $87.5\%$ mean pass@1, compared with 29 of 32 cases and $75.0\%$ for \textit{R3A}.
These results suggest the potential of combining generative search with constrained symbolic repair, although the cross-study comparison does not isolate the contribution of SMT reasoning and both methods depend on the adequacy of the waveform checks.

Agentic workflows can also target debugging outputs other than RTL repair.
\textit{MARVEL}~\cite{dbg_marvel} coordinates a security-policy supervisor with tool-specialized executors to identify SoC vulnerabilities rather than generate patches.
Its 51 reported findings comprise 19 vulnerabilities, 14 warnings, and 18 hallucinated findings, highlighting both the coverage and false-positive challenges of agentic security analysis.
\textit{AssertFix}~\cite{dbg_assertfix} targets assertion repair by retrieving relevant RTL and distinguishing timing from logic errors before correction.
It achieves formal assertion-fix rates of $73.7\%--83.8\%$ across four designs, but assumes golden RTL and therefore addresses a different repair target from faulty-RTL correction.

\textit{LintLLM}~\cite{dbg_lintllm} illustrates structured diagnostic reasoning without an explicitly established multi-agent architecture.
Its Logic-Tree prompt decomposes diagnosis into ordered steps, while a Defect Tracker tentatively fixes detected defects and reruns LLM-based detection to identify primary defects.
Because this feedback comes from repeated LLM detection rather than independent EDA-tool validation, the resulting diagnoses still require external confirmation.
Across 90 designs spanning 11 defect categories and three difficulty levels, it reports an $83.33\%$ correct rate and a $12.22\%$ false-positive rate.

Overall, structured and agentic methods complement the knowledge-oriented mechanisms of the preceding categories by explicitly controlling how debugging evidence is represented, analyzed, and searched.
Dependency graphs can expose spatial and temporal fault propagation, structured search can explore alternative repairs, and specialized agents or symbolic tools can decompose and constrain complex debugging tasks.
These mechanisms are particularly useful when failures cannot be resolved from a single prompt or local code context, but their effectiveness remains bounded by the quality of the available verification evidence, reference models, and tool feedback.
In practice, the four categories are complementary rather than exclusive: hybrid frameworks increasingly combine prompting, retrieval, model adaptation, structured reasoning, and verification tools to address different information and control requirements throughout the debugging process.
\begin{table*}[!t]
\caption{Summary of LLM-based hardware debugging works (Part I). Cat.: C1 = prompting/model generation/tool feedback, C2 = retrieval, C3 = parameter adaptation, C4 = structured/agentic reasoning. BL = bug localization, RCA = root cause analysis, Fix = RTL repair, E2E = integrated failure-to-patch workflow with tool rechecking.}
\label{tab:debug-part1}
\centering
\begingroup
\fontsize{7.5}{8.7}\selectfont
\setlength{\tabcolsep}{1.8pt}
\renewcommand{\arraystretch}{1.15}
\begin{tabularx}{\textwidth}{@{}>{\raggedright\arraybackslash}p{.12\textwidth}>{\centering\arraybackslash}p{.04\textwidth}>{\centering\arraybackslash}p{.03\textwidth}>{\centering\arraybackslash}p{.033\textwidth}>{\centering\arraybackslash}p{.03\textwidth}>{\centering\arraybackslash}p{.038\textwidth}>{\raggedright\arraybackslash}X>{\raggedright\arraybackslash}p{.075\textwidth}>{\raggedright\arraybackslash}p{.185\textwidth}>{\raggedright\arraybackslash}p{.18\textwidth}@{}}
\hline
\textbf{Paper} & \textbf{Cat.} & \textbf{BL} & \textbf{RCA} & \textbf{Fix} & \textbf{E2E} & \textbf{Core Technique} & \textbf{Bug Types} & \textbf{Key Result} & \textbf{Benchmark} \\
\hline
\textit{AutoVeriFix}~\cite{dbg_autoverifix} & C1 & -- & -- & $\checkmark$ & $\checkmark$ & Python-assisted two-stage generation, iterative simulation feedback & Syntax, functional & GPT-4: 77.2\% pass@1 (Human), 83.7\% pass@1 (Machine), 83.5\% pass@5 (RTLLM v2.0) & VerilogEval-Human/Machine, RTLLM v2.0, generation outcomes \\
\hline
\textit{VerilogCoder}~\cite{dbg_verilogcoder} & C1,4 & $(\checkmark)$ & $(\checkmark)$ & $\checkmark$ & $\checkmark$ & Graph-based planning, debug agent, AST-based waveform tracing & Syntax, functional & GPT-4 Turbo: 94.2\% workflow pass rate with supplied testbench feedback & VerilogEval-Human v2, 156 tasks, supplied golden testbenches \\
\hline
\textit{On hardware security bug code fixes}~\cite{dbg_securityfix} & C1 & -- & -- & $\checkmark$ & $(\checkmark)$ & CWE-conditioned prompts, model/prompt sampling, selected CWEAT integration & Hardware security & 15/15 cases have a successful candidate, GPT-4: 3,862/7,500 correct attempts & 15 security bugs, separate functional-repair comparison \\
\hline
\textit{MEIC}~\cite{dbg_meic} & C1,4 & $(\checkmark)$ & $(\checkmark)$ & $\checkmark$ & $\checkmark$ & Prompt-configured debug/scorer agents, candidate history and rollback & Syntax, functional & 93\% syntax and 78\% functional fix rates & 178 RTLLM-derived error instances \\
\hline
\textit{UVLLM}~\cite{dbg_uvllm} & C1,4 & $\checkmark$ & $(\checkmark)$ & $\checkmark$ & $\checkmark$ & UVM checking, mismatch-driven DFG slicing, test-score rollback & Syntax, functional & 86.99\% syntax and 71.92\% functional fix rates & 331 error instances across 27 RTLLM-derived modules \\
\hline
\textit{RTLFixer}~\cite{dbg_rtlfixer} & C1,2,4 & -- & $(\checkmark)$ & $\checkmark$ & $\checkmark$ & ReAct and expert-guidance retrieval with compiler feedback & Compilation errors & 98.5\% compilation-error repair, not a functional repair rate & VerilogEval-derived syntax errors, additional RTLLM evaluation \\
\hline
\textit{From Bugs to Fixes}~\cite{dbg_bugsfixes} & C2 & $\checkmark$ & $(\checkmark)$ & $\checkmark$ & -- & Specification retrieval, expression decomposition and patching closure & Functional mutations & 25/25 nonconstant mutations patched, 0/5 constant mutations & AON Timer, UART, and EDN, 30 injected mutations \\
\hline
\textit{GROVE}~\cite{dbg_grove} & C2,4 & $(\checkmark)$ & $(\checkmark)$ & $\checkmark$ & -- & Tool-validated knowledge tree, applicability conditions, snapshot-and-zoom retrieval & Assertion-triggering RTL faults & Llama-3.3: 84.7\% pass@1, 86.9\% pass@5, Jasper-checked & SVA-Eval: 732 training/183 test cases, module-disjoint split \\
\hline
\textit{HDLdebugger}~\cite{dbg_hdldebugger} & C2,3 & $(\checkmark)$ & $(\checkmark)$ & $\checkmark$ & -- & Reverse-engineered error pairs, code/document retrieval, supervised fine-tuning & Compiler-oriented HDL errors & 81.93\% execution pass rate on the reported test split & 92,143 industrial HDL instances, 8:2 train/test split \\
\hline
\textit{HDLxGraph}~\cite{dbg_hdlxgraph} & C2 & $(\checkmark)$ & -- & $(\checkmark)$ & -- & AST hierarchy and DFG signal traversal for repository retrieval & Repository defects & Debugging assessed by ROUGE overlap, no functional repair pass rate & Five \texttt{mor1kx} commits, HDLSearch evaluates retrieval separately \\
\hline
\textit{ViTAD}~\cite{dbg_vitad} & C2,4 & $\checkmark$ & $\checkmark$ & $\checkmark$ & -- & Timing dependency graph, violation-path analysis, cause-conditioned repair knowledge & Setup/hold, CDC & GPT-4o: 71.05\% fix rate versus 54.38\% without \textit{ViTAD} & 54 reported injected timing cases, Vivado-based evaluation \\
\hline
\textit{LLM4SecHW}~\cite{dbg_llm4sechw} & C3 & $\checkmark$ & $(\checkmark)$ & $\checkmark$ & -- & Repository-history defect/fix pairs and descriptions, adapter-based fine-tuning & Hardware defects, security & Per-case ROUGE and qualitative examples, no aggregate validated repair rate & Open-source hardware version histories, including OpenTitan \\
\hline
\end{tabularx}
\par\vspace{3pt}
\begin{minipage}{\textwidth}
\raggedright
$\checkmark$: explicitly implemented target/output, $(\checkmark)$: supporting operation or restricted demonstration, --: not separately established. E2E denotes automated tool rechecking within the declared workflow, not full-design verification, signoff, or independent proof of all unaffected behavior. An E2E dash does not mean that a paper omitted experimental validation. Category labels identify mechanisms, not publication order. Metrics retain each paper's evaluation definition and are not directly comparable across rows.
\end{minipage}
\endgroup
\end{table*} 

\begin{table*}[!t]
\caption{Summary of LLM-based hardware debugging works (Part II). The column definitions follow Table~\ref{tab:debug-part1}. SO denotes source-only screening as an input condition, not an additional category. A superscript A identifies assertion repair rather than RTL repair.}
\label{tab:debug-part2}
\centering
\begingroup
\fontsize{7.5}{8.7}\selectfont
\setlength{\tabcolsep}{1.8pt}
\renewcommand{\arraystretch}{1.15}
\begin{tabularx}{\textwidth}{@{}>{\raggedright\arraybackslash}p{.12\textwidth}>{\centering\arraybackslash}p{.04\textwidth}>{\centering\arraybackslash}p{.03\textwidth}>{\centering\arraybackslash}p{.033\textwidth}>{\centering\arraybackslash}p{.03\textwidth}>{\centering\arraybackslash}p{.038\textwidth}>{\raggedright\arraybackslash}X>{\raggedright\arraybackslash}p{.075\textwidth}>{\raggedright\arraybackslash}p{.185\textwidth}>{\raggedright\arraybackslash}p{.18\textwidth}@{}}
\hline
\textbf{Paper} & \textbf{Cat.} & \textbf{BL} & \textbf{RCA} & \textbf{Fix} & \textbf{E2E} & \textbf{Core Technique} & \textbf{Bug Types} & \textbf{Key Result} & \textbf{Benchmark} \\
\hline
\textit{LiK}~\cite{dbg_lik} & C3 & $\checkmark$ & -- & -- & -- & Continual pretraining, line-localization supervision, SimPO & Functional RTL faults & 93.38\% localization pass@1, downstream GPT-3.5 repair: 40.39\% to 58.92\% & 102 RTLLM-derived cases spanning five injected bug types \\
\hline
\textit{VeriDebug}~\cite{dbg_veridebug} & C3 & $\checkmark$ & -- & $\checkmark$ & -- & Contrastive localization/type embeddings and guided repair with shared parameters & Functional RTL faults & 67.2\% location Acc@1, 64.7\% exact-match repair Acc@1 & Synthetic functional defects, compilable RTL, held-out evaluation split \\
\hline
\textit{AssertSolver}~\cite{dbg_assertsolver} & C3 & $\checkmark$ & $(\checkmark)$ & $\checkmark$ & -- & Pretraining, supervised fine-tuning, and DPO on correct/incorrect repair responses & Assertion-triggering RTL faults & 88.54\% pass@1, 90.00\% pass@5, assertion-failure resolution & SVA-Eval: 877 generated and 38 human-curated cases \\
\hline
\textit{Self-HWDebug}~\cite{dbg_selfhwdebug} & C1 & -- & -- & $\checkmark$ & -- & Instructions synthesized from vulnerable/secure pairs, no weight update & Hardware security & Two-shot setting: 25/25 test snippets mitigated & Five CWE categories, five test snippets per category \\
\hline
\textit{FVDebug}~\cite{dbg_fvdebug} & C2,4 & $\checkmark$ & $\checkmark$ & $\checkmark$ & $(\checkmark)$ & Signal-event causal graph, balanced scanning, hypothesis exploration, patch generation & Formal-verification failures & Quality@Best 0.956 (LLM judge), repair pass@1/pass@5: 71.1\%/86.8\% & 38 SVA-Eval-Human cases, additional CVA6 and industrial RCA cases \\
\hline
\textit{R3A}~\cite{dbg_r3a} & C4 & $\checkmark$ & $(\checkmark)$ & $\checkmark$ & $\checkmark$ & AST-segmented localization agents, stochastic tree-of-thoughts patch search & Structural, functional & 29/32 cases solved, mean pass@1 75.0\%, pass@5 86.7\% & RTL-Repair, acceptance against reference waveforms \\
\hline
\textit{LintLLM}~\cite{dbg_lintllm} & C4, SO & $\checkmark$ & $(\checkmark)$ & $(\checkmark)$ & -- & Logic-Tree prompting, tentative fixes and LLM redetection to isolate primary defects & Syntax, lint/semantic defects & o1-mini: 83.33\% correct rate, 12.22\% reported false-positive rate & 90 designs, 11 defect categories, three difficulty levels \\
\hline
\textit{FLAG}~\cite{dbg_flag} & C1, SO & $\checkmark$ & -- & -- & -- & Static targeting, regeneration differences and token probabilities, line classifier & Security, functional & Line-level detection: 51/201, token top-five: 47/201 & OpenTitan (120), HackAtDAC (42), CirFix (30), CWE examples (9) \\
\hline
\textit{ARSP}~\cite{dbg_arsp} & C3,4 & $(\checkmark)$ & -- & $\checkmark$ & -- & Trained semantic partitioning and fragment repair, controlled recombination & Functional RTL faults & 77.92\% pass@1, 83.88\% pass@5, testbench-checked & Injected faults in 33 RTLLM/GenBen designs, mean 219 RTL lines \\
\hline
\textit{VeriPilot}~\cite{dbg_veripilot} & C4 & $\checkmark$ & $(\checkmark)$ & $\checkmark$ & $\checkmark$ & Golden-model variable alignment, CDFG tracing, reference-guided iterative repair & Functional RTL faults & GPT-4o on adapted CVDP: 54.3\% to 85.71\% repair success & Strider and adapted CVDP tasks, executable golden models supplied \\
\hline
\textit{Clover}~\cite{dbg_clover} & C4 & $(\checkmark)$ & $(\checkmark)$ & $\checkmark$ & $\checkmark$ & Agentic dispatch to LLM/SMT repair, dialogue-state tree search, RTL toolbox & Structural, functional & 31/32 RTL-Repair cases solved, mean pass@1 87.5\% & RTL-Repair, additional random-logic synthetic evaluation \\
\hline
\textit{MARVEL}~\cite{dbg_marvel} & C2,4 & $\checkmark$ & $(\checkmark)$ & -- & -- & Security-policy supervisor and tool-specialized detection executors & SoC security & 51 reports: 19 vulnerabilities, 14 warnings, 18 hallucinations & OpenTitan-based Hack@DATE SoC \\
\hline
\textit{AssertFix}~\cite{dbg_assertfix} & C2,4 & $(\checkmark)$ & $\checkmark$ & $\checkmark^{\mathrm{A}}$ & $(\checkmark)^{\mathrm{A}}$ & Related-RTL retrieval, timing/logic classification, assertion-specific correction & Incorrect SVA timing/logic & Formal assertion fix rates: 73.7\%--83.8\% across four designs & I2C, ECG, Pairing, SHA3, golden RTL assumed \\
\hline
\end{tabularx}
\par\vspace{3pt}
\begin{minipage}{\textwidth}
\raggedright
BL includes candidate-region extraction. It does not imply measured line-level accuracy. RCA marked as supporting denotes explanatory reasoning or defect classification without an independently evaluated causal-diagnosis task. \textit{LintLLM}'s edits support internal diagnosis. \textit{FVDebug}'s large-design demonstrations concern RCA and should not inherit its small-benchmark repair rates. Assertion repair and source-only detection are included for their debugging role but remain distinct from RTL functional repair.
\end{minipage}
\endgroup
\end{table*}

\subsection{LLM-Based Debugging Works Across Pipeline Stages}
\label{subsec:debug-pipeline}

The preceding taxonomy classifies LLM-based debugging methods by their primary mechanisms.
We next compare the reviewed works by their coverage of the debugging pipeline in Fig.~\ref{fig:debug_process}, focusing on bug localization, RCA, repair, and integrated re-verification.
Tables~\ref{tab:debug-part1} and~\ref{tab:debug-part2} summarize these capabilities and the corresponding evaluation evidence.

\textbf{Bug localization.}
Localization methods differ in both the granularity of their outputs and how directly those outputs identify faulty RTL.
\textit{LiK}~\cite{dbg_lik} predicts fault locations that guide downstream correction, while \textit{VeriDebug}~\cite{dbg_veridebug} jointly learns localization and repair.
In contrast, \textit{ARSP}~\cite{dbg_arsp} identifies a candidate repair region, and \textit{HDLxGraph}~\cite{dbg_hdlxgraph} retrieves repository context relevant to diagnosis.
These outputs serve related purposes but provide different levels of localization evidence.
Notably, \textit{LiK} shows that its predicted locations improve downstream GPT-3.5 repair from 40.39\% to 58.92\%, demonstrating practical value beyond localization accuracy.
By comparison, candidate-region extraction or context retrieval alone does not establish line-level localization accuracy or improved repair.
Localization should therefore be evaluated by both the specificity of the identified region and its demonstrated contribution to subsequent debugging.

\textbf{Root cause analysis.}
RCA goes beyond localization by explaining how the suspected RTL produces the observed failure.
Its evaluation should therefore distinguish a causally supported diagnosis from reasoning generated only as an intermediate step toward repair.
\textit{FVDebug}~\cite{dbg_fvdebug} separately evaluates generated causal hypotheses and checked repair outcomes, reporting a Quality@Best score of 0.956 for RCA in addition to its repair results.
\textit{AssertSolver}~\cite{dbg_assertsolver} also generates explanations with proposed corrections, but explanation generation alone does not establish causal accuracy.
Similarly, a successful patch validates the correction only under the supplied checks and does not necessarily validate every step of its accompanying explanation.
Evaluation scale must also be associated with the pipeline stage: \textit{FVDebug}'s larger CVA6 and industrial demonstrations establish RCA capability at that scale, but not equivalent automated repair performance.
RCA evaluation should therefore report whether causal hypotheses are independently assessed and distinguish such evidence from intermediate reasoning used to guide repair.

\textbf{Repair and closed-loop debugging.}
Repair methods differ in both the artifact being corrected and the criterion used to accept a candidate.
For RTL repair, \textit{RTLFixer}~\cite{dbg_rtlfixer} evaluates compilation success, whereas \textit{VeriDebug}~\cite{dbg_veridebug} measures exact agreement with a reference corrected line.
These criteria differ from behavioral validation through simulation, assertions, or formal checking, as used by other repair frameworks~\cite{dbg_uvllm,dbg_fvdebug}.
\textit{AssertFix}~\cite{dbg_assertfix} further targets incorrect assertions while assuming golden RTL, and therefore addresses a different repair problem from correcting a faulty implementation.
Closed-loop integration adds another dimension by automatically rechecking candidate repairs and using the resulting feedback for further correction.
For example, \textit{UVLLM}~\cite{dbg_uvllm} combines candidate generation with UVM-based rechecking to support iterative repair.
Such re-verification strengthens the debugging loop but establishes correctness only with respect to the checks exercised by the workflow.

Overall, the reviewed methods provide increasingly broad coverage of the debugging pipeline, ranging from source-level localization to integrated localization, diagnosis, repair, and re-verification.
However, pipeline coverage alone does not determine debugging capability or correctness.
Localization outputs vary in granularity, RCA requires evidence that generated explanations reflect the observed failure, and repair results depend strongly on the acceptance criterion and behavioral oracle.
Accordingly, comparisons should distinguish \emph{which pipeline stages are automated} from \emph{what evidence validates the output of each stage}.
This distinction is particularly important for closed-loop workflows, where repeated tool feedback can improve repair success without guaranteeing correctness beyond the supplied verification checks.
\subsection{Debugging as a Component of LLM-Based RTL Generation Framework}
\label{subsec:debug-generation}

The preceding subsections consider debugging primarily as the diagnosis and repair of faulty RTL.
Debugging also serves as an integral component of LLM-based RTL generation, where generated candidates are iteratively corrected before being accepted.
In this setting, correction occurs at two levels: compilation errors can be addressed using compiler diagnostics, whereas functional errors require behavioral feedback from a testbench or reference model.

\textit{RTLFixer}~\cite{dbg_rtlfixer} illustrates compilation-oriented correction of LLM-generated Verilog.
It combines compiler feedback with retrieved expert repair guidance to iteratively revise erroneous candidates.
This improves RTL executability, but compilation alone does not establish functional correctness.

Functional correction requires stronger behavioral feedback.
\textit{VerilogCoder}~\cite{dbg_verilogcoder} integrates graph-based task planning with coding and debugging agents.
Its debugging agent uses simulation mismatches and AST-based waveform tracing to revise generated RTL against a supplied testbench.
With GPT-4 Turbo, the complete workflow achieves a 94.2\% pass rate on 156 VerilogEval-Human v2.
\textit{VeriTrace}~\cite{dbg_veritrace:iclad26} extends this paradigm by allowing a dedicated Inspector agent to iteratively select internal signals and temporal windows for hypothesis-driven waveform analysis before targeted RTL repair.

\textit{AutoVeriFix}~\cite{dbg_autoverifix} instead constructs its own behavioral reference from the natural-language specification.
Its first stage generates a Python reference model and tests, which are refined using Python coverage feedback.
The second stage generates Verilog and iteratively corrects compilation errors and behavioral mismatches against this reference.
With GPT-4, \textit{AutoVeriFix} achieves 77.2\% pass@1 on VerilogEval-Human, 83.7\% pass@1 on VerilogEval-Machine, and 83.5\% pass@5 on RTLLM v2.0.
These results characterize the complete generation-and-correction workflow rather than repair of a fixed set of pre-existing RTL defects.

Another recent work, \textit{VeriRefine}~\cite{kong2026progressive}, also integrates verification and debugging directly into RTL generation through a structured intermediate representation.
Unlike \textit{VerilogCoder}, which uses a supplied benchmark testbench to guide debugging, \textit{VeriRefine} generates its verification environment from the specification, keeping both generation and correction grounded in the same source of intended behavior.

Overall, debugging turns RTL generation into an iterative process guided by verification feedback. Its reliability depends on both the coverage of the checks and the correctness of the expected behavior. Verification artifacts generated from the specification may share interpretation errors with the RTL. Evaluation should therefore report feedback access and iteration budgets and use benchmark simulation or properties withheld during repair to assess functional correctness.
\subsection{Future Research Direction: Self-Evolving Hardware Debugging Systems}
\label{subsec:self-evolving-debug}

The reviewed methods demonstrate a progression from prompt-based correction toward retrieval, model adaptation, structured reasoning, agentic search, and verification-guided repair.
However, most systems learn little from completed debugging tasks and restart subsequent tasks with the same knowledge and strategies.
A promising direction is \emph{self-evolving hardware debugging}, in which agents accumulate validated experience across designs and revisions and use it to improve bug localization, diagnosis, repair, and verification.

A self-evolving system should retain more than successful patches.
Each debugging episode should associate the failure and repair with its fault location, causal evidence, relevant signal dependencies, specification assumptions, applicability conditions, and validation results.
\textit{GROVE}~\cite{dbg_grove} provides an initial example of organizing validated repair knowledge for reuse, while \textit{MEIC}, \textit{R3A}, and \textit{Clover}~\cite{dbg_meic,dbg_r3a,dbg_clover} maintain candidate states and feedback within individual debugging sessions.
Extending such mechanisms across tasks could allow previous experience to guide retrieval, fault localization, repair selection, and tool invocation rather than merely provide another repair example.

A central challenge is \emph{reliable transfer across designs and revisions}.
A repair that succeeds on one design may become invalid when interfaces, timing behavior, implementation structure, or specifications change.
Persistent knowledge must therefore retain the conditions under which it was learned and be revalidated before reuse rather than treating every successful repair as universally applicable.
Evaluation should correspondingly use sequences of unseen designs or design revisions and compare systems with and without accumulated experience under equivalent resource budgets.
Such experiments should measure whether experience reduces localization effort and improves independently validated repair rates without increasing regressions or false positives.

A second challenge is \emph{evolving debugging strategies rather than only accumulating repair knowledge}.
The structured and agentic methods reviewed above show that debugging performance depends on decisions about which fault hypotheses to explore, which signal or temporal dependencies to trace, which context to retrieve, and when to invoke simulation, formal, or symbolic tools.
Future agents could adapt these policies from previous validated outcomes, progressively learning which reasoning and tool-use strategies are effective for different failure classes.
Such adaptation must be separated experimentally from gains due simply to additional retrieval, repair attempts, or computation.
Runtime, token consumption, tool calls, and solver budgets should therefore accompany repair accuracy when evaluating self-evolving strategies.

A third challenge is maintaining \emph{independent and trustworthy verification}.
The preceding studies show that compiler success, exact patch matching, simulation tests, assertion checking, and formal verification provide different levels of evidence, and no single metric establishes correctness in all settings.
Moreover, when the same testbench or reference model guides repair and determines acceptance, the system can overfit to the available verification environment.
A self-evolving agent could amplify this problem by retaining such locally successful but incomplete repairs as future knowledge.
New experience should therefore be admitted to persistent memory only after independent validation, preferably using held-out tests, properties, or formal checks that are not available during repair.
The agent should also distinguish faults in the RTL from errors in assertions, reference models, specifications, or environmental assumptions rather than modifying the artifact that defines its own acceptance criterion.

Finally, scalability remains an open challenge because realistic failures can span modules, clock cycles, and long execution traces.
Current approaches demonstrate the benefits of dependency graphs, slicing, retrieval, fine-tuning, and agent decomposition, but evidence for large-scale end-to-end repair remains limited.
Future systems will require compact representations of spatial and temporal failure evidence, hierarchical localization and reasoning, selective retrieval, and adaptive tool invocation to control context and computational cost.
Human oversight will also remain important for ambiguous specifications, high-impact repairs, and final acceptance criteria.

Overall, existing work establishes that LLMs can support individual stages of hardware debugging and that combining prompting, retrieval, adaptation, structured reasoning, and verification feedback can substantially strengthen closed-loop repair.
The next step is to move from isolated debugging episodes toward systems that \emph{learn from validated experience while preserving independent verification}.
Achieving this goal requires reliable cross-design transfer, persistent knowledge with explicit applicability conditions, adaptive yet resource-efficient debugging policies, scalable reasoning over temporal and hierarchical RTL behavior, and evaluation against verification evidence withheld from the learning and repair process.
Addressing these challenges is essential for transforming LLM-based debugging from task-specific repair automation into dependable hardware debugging systems that improve across designs without accumulating incorrect or obsolete knowledge.
\section{LLMs for Formal Verification}
\label{sec:llm4fv}

Simulation-based validation remains indispensable in hardware verification because of its flexibility and scalability; however, it explores only a finite set of input sequences and execution scenarios and therefore cannot guarantee the absence of bugs. Formal verification (FV), in contrast, provides exhaustive mathematical guarantees by proving that a design satisfies specified properties over all states and inputs admitted by a formal model \cite{clarke1999model,biere2003bmc}. Despite these guarantees, practical adoption of FV is hindered by two fundamental bottlenecks. The first is the specification bottleneck, which requires engineers to translate design intent into precise properties, environmental assumptions, invariants, and verification models. The second is the computational bottleneck arising from state-space explosion and the complexity of SAT/SMT-based reasoning for large designs. Industrial FV further requires engineers to select appropriate proof engines and strategies, analyze counterexamples, diagnose root causes, and iteratively refine verification models. These challenges are compounded by incomplete or ambiguous specifications, bounded-verification limitations, vacuous properties, multi-property interactions, and solver scalability, making formal verification both labor intensive and computationally demanding.

Recent advances in large language models (LLMs) have created opportunities to address both the specification and computational bottlenecks of formal verification. By processing natural-language specifications, RTL, assertions, source code, and verification traces, LLMs can generate candidate properties, invariants, verification models, proof strategies, and solver optimization heuristics, while also assisting with counterexample analysis and verification workflow orchestration. Rather than relying directly on LLM outputs, recent research increasingly adopts a neuro-symbolic paradigm in which LLMs perform semantic reasoning and heuristic generation, while model checkers, SAT/SMT solvers, and other formal engines validate the generated artifacts and provide corrective feedback. This separation combines the flexibility of generative models with the soundness of formal verification and has enabled applications spanning property generation, model checking, logic equivalence checking, solver optimization, and end-to-end verification workflow automation.

SystemVerilog assertion generation using LLMs has been treated in Section~\ref{sec:sva-gen}. This section reviews the emerging landscape of LLM-driven formal verification and examines how LLMs are being integrated with formal engines to address proof construction, equivalence checking, solver optimization, and workflow automation. It also analyzes the limitations that currently constrain reliability and scalability, including hallucination, limited context and reasoning depth, data scarcity, benchmark overfitting and contamination, verification latency, and solver scalability. By organizing existing approaches according to their verification objectives, LLM roles, formal back ends, and evaluation methodologies, this review identifies common design patterns and limitations and highlights research directions toward reliable, scalable, and increasingly autonomous LLM-assisted formal verification for production-grade hardware.

\subsection{LLM-Assisted Proof Generation and Model Checking}

Beyond formal property generation, recent research has increasingly employed LLMs to assist the proof process itself. These approaches use LLMs to construct intermediate proof artifacts, guide proof search, optimize model-checking algorithms, and coordinate verification workflows, thereby moving LLMs deeper into the formal verification pipeline while retaining conventional formal engines as the source of correctness guarantees.

One important direction is incremental and hierarchical proof generation, which decomposes large verification problems into smaller proof obligations. Qayyum \emph{et al.}~\cite{qayyum2026llm} proposed an LLM-assisted framework that partitions a design into smaller modules and generates module-level invariants and interface properties. Each generated property is independently validated using SAT/SMT-based formal verification before being composed into higher-level proofs. This decomposition reduces the reasoning burden on the LLM and improves proof scalability by avoiding direct reasoning over the complete flattened design.

A closely related direction is LLM-assisted invariant synthesis, where the LLM generates candidate inductive invariants that are subsequently validated by formal engines. Hassan \emph{et al.}~\cite{Hassan2024Mutation} combined GPT-4-generated invariants with formal verification and mutation testing. Mutation testing provides a stronger assessment of invariant quality than syntactic validity or proof success alone because it evaluates whether generated invariants can distinguish the original design from intentionally modified implementations, thereby exposing semantically weak or vacuous properties. CIll~\cite{Su2026CIll} further introduced a counterexample-to-induction (CTI)-guided approach in which candidate invariants are checked using bounded model checking and inductiveness checking, and counterexamples are returned to the LLM to guide iterative refinement. Integration with IC3 and $k$-induction enabled verification of multiple RISC-V processor designs. Together, these studies demonstrate the potential of closed-loop LLM--formal-engine interactions for synthesizing nontrivial proof artifacts, while also showing that effectiveness depends strongly on the quality of counterexample interpretation and the efficiency of formal feedback.

A complementary direction applies LLMs not to individual proof artifacts but to the model-checking algorithms themselves. IC3-Evolve~\cite{Miao2026IC3Evolve}, for example, uses an LLM offline to evolve IC3/PDR heuristics through auditable code modifications. Candidate changes are accepted only when SAFE executions produce independently verifiable proof certificates and UNSAFE executions generate replayable counterexamples. The resulting optimized model checker operates without an LLM at runtime, separating heuristic discovery from trusted proof reasoning. This approach illustrates an important distinction between LLM-assisted verification, where the LLM participates in an individual verification instance, and LLM-driven verification optimization, where the LLM searches for improvements to the underlying formal engine.

Recent work has further expanded the role of LLMs from assisting individual verification tasks to orchestrating complete formal verification workflows. Saarthi~\cite{kumar2025saarthi}, for example, organizes specification analysis, verification planning, assertion generation, property proving, counterexample analysis, and coverage closure into an integrated multi-agent workflow coupled with commercial formal verification tools. Although its reported end-to-end automation rate remains approximately $40\%$, the framework demonstrates a transition toward agentic formal verification in which multiple reasoning and verification activities are coordinated within a unified workflow.

Collectively, these studies indicate an evolution from LLM-based generation of individual proof artifacts toward closed-loop proof refinement, model-checker optimization, and multi-agent verification orchestration. The current state of practice remains fundamentally neuro-symbolic: LLMs are effective as generators, heuristic search mechanisms, decomposition agents, and counterexample interpreters, whereas SAT/SMT solvers, bounded model checkers, IC3/PDR, and commercial formal engines remain responsible for establishing correctness. This division of responsibility is essential because linguistic plausibility cannot establish the soundness or completeness required for formal hardware verification.

\subsection{LLM-Assisted Logic Equivalence Checking}

Compared with LLM-assisted property generation and model checking, the use of LLMs for logic equivalence checking (LEC) remains relatively immature. LEC determines whether two hardware implementations exhibit equivalent behavior and is routinely used for RTL-to-RTL comparison, RTL-to-gate verification, and validation of synthesis and engineering-change-order (ECO) transformations. Because conventional LEC provides a proof or counterexample rather than a statistical confidence estimate, current LLM-based approaches generally use LLMs in conjunction with formal equivalence engines rather than as replacements for them. Two complementary directions have emerged: learning-based prediction of equivalence relationships and the use of formal equivalence checking as a correctness oracle for LLM-based RTL generation and refinement.

DeepRTL2~\cite{liu2025deeprtl2} explores RTL equivalence as a representation learning problem. It constructs equivalent and non-equivalent RTL pairs through a feedback-driven process in which GPT-4o generates functionality-preserving RTL transformations and a conventional Yosys LEC flow determines whether the resulting designs remain equivalent. The generated pairs are subsequently used to train representations for equivalence classification. DeepRTL2 reports $68.1\%$ accuracy and $72.3\%$ F1 score, outperforming the evaluated general-purpose embedding baselines. However, the learned predictor serves primarily as a filtering mechanism rather than a replacement for formal equivalence checking because statistical classification cannot provide exhaustive correctness guarantees.

A more prevalent approach treats formal equivalence checking as a correctness oracle within LLM-driven RTL generation. FormalEval~\cite{formaleval} uses sequential equivalence checking to compare generated RTL with a reference implementation and returns either a proof result or a counterexample. The formal result therefore provides an objective functional-correctness signal for evaluating and refining LLM-generated RTL. FormalRTL~\cite{formalrtl2026} extends this paradigm by coupling LLM-based planning and RTL synthesis with formal equivalence checking and software reference models as executable specifications. Generated RTL is accepted or revised according to formal feedback, transforming RTL generation from a one-shot synthesis task into an iterative generate--verify--repair process.

These studies reveal a common pattern between LLM-assisted model checking and LEC: in both settings, LLMs primarily perform semantic reasoning and generate or refine verification artifacts, while formal engines provide objective feedback that guides subsequent iterations. The key distinction lies in the verification objective: model-checking approaches use LLMs to construct invariants, decompose proof obligations, interpret counterexamples, or guide proof search for properties of a design, whereas LEC approaches use LLMs to reason about correspondences between implementations and to synthesize or repair RTL under equivalence constraints. 
On the other hand, this common \emph{generate--formally validate--refine} paradigm suggests that techniques developed for LLM-guided proof refinement, counterexample interpretation, and hierarchical decomposition could potentially be transferred to LEC, particularly for correspondence identification and equivalence-problem partitioning.
It is also interesting to extend the role of LLMs toward broader proof assistance, including identification of correspondence points, abstraction and partitioning strategies, and interpretation of equivalence counterexamples.

\subsection{LLM-Driven SAT/SMT Solver Optimization}

Modern formal verification engines, including IC3/PDR-based model checkers, rely heavily on SAT/SMT solvers as their core reasoning engines. Their performance is therefore strongly influenced by solver heuristics and implementation choices, including branching, clause learning and deletion, preprocessing and inprocessing, restarts, and theory propagation. This dependency has motivated research on using LLMs to automate the discovery and optimization of solver heuristics and implementations.

Non-LLM solver specialization provides an important baseline for this research direction. Su \emph{et al.} developed GipSAT~\cite{su-gipSAT:2025}, a lightweight CDCL solver specialized for IC3 workloads by exploiting structural properties of IC3-generated SAT queries. GipSAT reduces unnecessary decisions, replaces the conventional VSIDS binary heap with a bucket-based data structure, and optimizes temporary-clause management, achieving an average $3.61\times$ speedup over a MiniSat-based IC3 implementation on HWMCC'20 and HWMCC'24 benchmarks. The subsequent rIC3 framework further combines SAT-level optimizations with IC3 improvements such as dynamic generalization and internal-signal-based reasoning~\cite{su-ric3:2025}. These results establish that substantial gains can arise from tailoring both solver and model-checking strategies to verification-specific workloads and motivate LLMs as mechanisms for automating such optimization.

Based on these observations, AutoSAT~\cite{autosat:sun2026} and SolSearch~\cite{solsearch:sheng2025} use LLMs to generate and iteratively refine CDCL solver heuristics or source-code modifications based on benchmark performance. AutoModSAT~\cite{autoModSat:sun2025discovering} further combines prompt optimization with evolutionary search to explore modular SAT solver configurations. These approaches formulate solver optimization as an empirical search loop in which the LLM proposes a modification, the resulting solver is compiled and evaluated on representative instances, and performance feedback guides subsequent iterations. The results demonstrate that LLMs can explore a broad design space of solver heuristics and implementation strategies that would otherwise require substantial manual expertise.

LLM-based optimization has also been explored for SMT reasoning. LLMs can generate and refine SMT constraints or solver API code from high-level objectives, translating informal requirements into SMT-LIB formulations or APIs such as Z3 Optimize~\cite{hwang2026agentic}. For solver optimization, however, syntactic validity is insufficient: generated formulations and solver modifications must preserve semantic correctness while improving solving efficiency and avoiding unsupported or computationally inefficient theory combinations. Consequently, the central challenge is to jointly evaluate correctness and performance rather than optimize either criterion in isolation.

Overall, existing work have demonstrated the search capability of LLMs for SAT/SMT solver optimization, particularly when candidate generation is combined with automated compilation, correctness checking, and empirical performance evaluation. 
However, current results remain vulnerable to benchmark overfitting, unstable optimization trajectories, compilation and execution failures, and loss of generality when optimizations are tuned to particular workloads. Future progress therefore requires evaluation across diverse formal-verification workloads and explicit measurement of both solver performance and generalization, rather than reporting improvements on a single benchmark distribution.

\subsection{Cross-Cutting Challenges and Future Directions}

Across these research directions, the state of the art is moving beyond one-shot LLM generation toward closed-loop, tool-grounded, and increasingly agentic formal verification. Current systems can generate assertions and invariants, decompose verification problems, interpret counterexamples, guide proof search, optimize solver implementations, and coordinate multi-stage verification workflows. The most reliable architectures share a common pattern: the LLM performs semantic reasoning, candidate generation, heuristic exploration, or workflow coordination, while deterministic formal tools provide correctness feedback and ultimately establish the validity of the generated artifacts. Nevertheless, current systems remain far from autonomous production-grade verification, and the primary research challenge is no longer simply whether LLMs can generate useful verification artifacts, but whether they can do so reliably, efficiently, scalably, and with sufficient generalization across designs and verification tasks.

First, correctness remains fundamentally different from linguistic plausibility. LLM-generated assertions, invariants, proof strategies, or solver modifications may appear reasonable while being incomplete, over-constrained, vacuous, or semantically incorrect. Formal validation is therefore essential, but validation alone does not guarantee useful verification if the generated properties do not adequately capture design intent. Future systems should jointly address property quality and proof validity through mechanisms that detect vacuity, measure specification coverage, identify missing behaviors, and use counterexamples to refine both the generated artifact and its underlying specification.

Second, scalability remains a central limitation at both the LLM and formal-engine levels. Large industrial designs exceed practical context limits and contain long-range dependencies that are difficult for LLMs to preserve across iterative interactions. At the same time, formal engines face state-space explosion and expensive SAT/SMT reasoning. Hierarchical verification, modular context management, retrieval of design-relevant information, and multi-agent decomposition are promising approaches, but their effectiveness must be evaluated against the cost of maintaining consistency across agents and abstraction boundaries. In particular, future systems should avoid merely shifting the scalability problem from the formal engine to an increasingly large LLM context.

Third, the interaction between LLMs and formal tools introduces a significant latency bottleneck. Proof attempts and SAT/SMT evaluations may require minutes or hours, making naive generate--verify--repair loops prohibitively expensive. Research should therefore investigate adaptive feedback mechanisms that determine when formal evaluation is necessary, surrogate models that prioritize promising candidates, and compact representations of counterexamples, proof failures, and solver traces that preserve the information most useful for subsequent LLM reasoning.

Fourth, generalization and evaluation remain insufficiently understood. Many existing studies rely on small benchmark suites or task-specific datasets, making it difficult to determine whether reported gains reflect genuine reasoning improvements or overfitting to particular designs, coding styles, solver configurations, or benchmark distributions. Future evaluations should include larger and more diverse hardware designs, unseen architectures, cross-benchmark testing, multiple LLMs, and controlled ablation studies that isolate the contributions of prompting, retrieval, agentic decomposition, formal feedback, and solver optimization. Evaluation should also report not only correctness and proof success, but verification time, tool-call cost, coverage, robustness across repeated runs, and the degree of human intervention required.

Finally, a major opportunity lies in developing hardware-aware agentic systems that jointly reason about specifications, RTL structure, formal properties, proof strategies, and the underlying SAT/SMT infrastructure. Rather than treating these tasks as independent stages, future frameworks could establish feedback loops in which failures at one level inform decisions at another—for example, counterexamples could guide property refinement, failed proofs could trigger abstraction or decomposition, and solver-performance feedback could influence model-checking strategies. Such systems should remain grounded by independently checkable proof artifacts and formal correctness oracles, enabling increasing levels of automation without sacrificing the soundness guarantees that distinguish formal verification from statistical software generation. 

\section{Conclusion and Future Directions}
\label{sec:conclusion}

This survey examined the emerging role of LLMs across hardware functional verification, including assertion generation, stimulus and testbench generation, bug localization and design repair, formal verification, equivalence checking, solver optimization, and agentic verification workflows.
Across these areas, a common pattern emerges: LLMs are most effective not as replacements for established verification algorithms, but as semantic reasoning and orchestration layers that connect specifications, design artifacts, verification evidence, and deterministic EDA tools.
Prompting, retrieval, model adaptation, structured representations, and agentic reasoning improve different aspects of this interaction, while simulators, formal engines, coverage tools, and SAT/SMT solvers remain essential for evaluating generated artifacts and establishing correctness.

A second cross-cutting finding is the growing importance of closed-loop, verification-grounded workflows.
Compilation diagnostics, simulation traces, coverage reports, counterexamples, proof outcomes, and solver statistics increasingly serve not only as acceptance signals but also as feedback for localization, refinement, search, and tool selection.
This feedback-driven paradigm appears across assertion and testbench generation, RTL debugging and repair, model checking, equivalence checking, and solver optimization.
However, successful interaction with a verification tool should not be confused with complete verification: an artifact may satisfy the available checks while still failing to capture behaviors or requirements that those checks do not represent.
The central challenge is therefore not merely generating executable or tool-accepted artifacts, but preserving alignment between \emph{design intent} and the \emph{verification evidence} used to justify correctness.

This observation motivates a broader research direction toward \emph{specification-centered verification intelligence}.
Future systems should maintain explicit relationships among requirements, RTL, assertions, stimuli, reference models, coverage objectives, counterexamples, and repair decisions rather than treating these artifacts independently.
Such traceability would enable an agent to determine whether a verification failure originates from the implementation, a generated verification artifact, an incorrect environmental assumption, or an incomplete specification.
It would also provide a stronger basis for measuring verification completeness and for preventing implementation-specific behavior from silently redefining the intended specification.
Structured intermediate representations, retrieval, multimodal reasoning, and hierarchical hardware knowledge are promising mechanisms for maintaining these relationships across large designs.

Scalability represents a second high-level challenge.
Current results are strongest on relatively localized tasks, whereas industrial verification requires reasoning across large RTL hierarchies, long temporal behaviors, heterogeneous verification artifacts, and expensive EDA-tool interactions.
Simply increasing context length, agent count, or refinement iterations is unlikely to provide a sustainable solution because it can shift rather than eliminate the underlying complexity.
Future systems should instead selectively retrieve and compress relevant information, reason hierarchically across abstraction levels, decompose verification objectives, and invoke expensive tools only when their expected information value justifies the cost.
The longer-term goal is an adaptive neuro-symbolic workflow in which LLM reasoning determines \emph{what} should be analyzed next while deterministic verification engines determine \emph{whether} the resulting claims are valid.

A third challenge is establishing generalization and trustworthy evaluation.
Reported results remain difficult to compare because studies differ substantially in benchmark scale, available design information, verification environments, model configurations, iteration budgets, and acceptance criteria.
Public hardware repositories further introduce potential overlap between model training data and evaluation artifacts.
Future benchmarks should therefore emphasize unseen and contamination-controlled designs, diverse architectures and abstraction levels, realistic specifications and defects, and evaluation across multiple models and verification tools.
Evaluation should measure not only final artifact success but also verification effectiveness, robustness across repeated runs, computational and tool cost, and required human intervention.
For iterative and agentic systems, controlled ablations are particularly important for distinguishing genuine improvements in reasoning from gains obtained through additional search, tool calls, or computation.

These considerations also change how autonomy should be viewed in hardware verification.
The emerging trajectory is from one-shot artifact generation toward persistent agents that plan verification activities, interpret heterogeneous evidence, coordinate tools, and reuse validated experience across designs and revisions.
Such systems could progressively improve retrieval, decomposition, debugging, and verification strategies, but persistent knowledge must retain its applicability conditions and validation evidence so that obsolete or design-specific conclusions are not propagated.
Human engineers will remain essential for resolving ambiguous intent, reviewing high-impact decisions, defining verification objectives, and establishing signoff criteria.
Accordingly, the objective should be dependable human--AI collaboration rather than unrestricted autonomous verification.

Overall, the reviewed literature shows that LLMs can substantially reduce the semantic and orchestration burden of hardware verification when coupled with deterministic verification infrastructure.
The field has progressed from isolated generation of assertions, tests, and repairs toward integrated workflows that reason over specifications, verification artifacts, failure evidence, and formal engines.
The next major step is to make these workflows \emph{specification-aligned, scalable, independently verifiable, and demonstrably generalizable}.
Progress along these dimensions could transform LLM-assisted verification from a collection of promising task-level demonstrations into a dependable verification methodology in which LLMs provide flexible reasoning and automation while established EDA tools and engineering judgment preserve the correctness guarantees required for hardware design.
\section*{Acknowledgment}

This work is supported by the National Science Foundation under Grant No. 2434247. 
Any opinions, findings, and conclusions or recommendations expressed in this material are those of the author(s) and do not necessarily reflect the views of the funding agencies.

\newpage
\addcontentsline{toc}{section}{References}
\bibliographystyle{ieeetr}
\bibliography{ref-misc,ref-fv,ref-rtl-gen,ref-tbgen,ref-svagen,ref-debug}
\end{document}